\documentstyle[eqsecnum,aps,psfig,epsf,12pt]{revtex}

\def\beq{\begin{equation}}
\def\eeq{\end{equation}}
 
\def\bea{\arraycolsep .1em \begin{eqnarray}}
\def\eea{\end{eqnarray}}

\let\be=\beta

\let\de=\delta
\let\eps=\epsilon
                  
\let\la=\lambda
                 
\let\si=\sigma
\let\La=\Lambda                 
\let\F=\varphi
\let\om=\omega

\let\Ga=\Gamma
\let\De=\Delta

\def\rb{\bar\rho}

\def\kcr{k_{\rm cr}}

\def\be2{\beta_{e^{2}}}         
\def\mst2{{\tilde m}_s^2}        

\def\ms2{m^2_s}
\def\Tr{{\rm Tr}}

\def\plusmal{\:\hbox to -0.8pt{\hbox{$\times$}\hss}{\hbox{$+$}}\:}

\def\e#1{e^#1}
\def\eb#1{\bar e^#1}
\def\es#1{e_{\star}^#1}

\def\t#1{\tilde #1}

\def\Es#1{eqs.~(\ref{#1})}
\def\Eq#1{eq.~(\ref{#1})}
\def\eq#1{(\ref{#1})}

\def\k-#1{k_{\rm #1}}

\def\s0#1#2{\mbox{\small{$ \frac{#1}{#2} $}}}
\def\0#1#2{\frac{#1}{#2}}

\def\el#1#2{\ell^{#1}_{#2}}

\def\6{\partial}

\def\grgl{\:\hbox to -0.2pt{\lower2.5pt\hbox{$\sim$}\hss}{\raise3pt\hbox{$>$}}\:}
\def\plusmal{\:\hbox to -0.2pt{\hbox{$\times$}\hss}{\hbox{$+$}}\:}
\def\klgl{\:\hbox to -0.2pt{\lower2.5pt\hbox{$\sim$}\hss}{\raise3pt\hbox{$<$}}\:}

\begin{document}
\thispagestyle{empty}
\begin{flushright}
{\tt HD-THEP/99-52\,\,\,\\
DIAS-STP-99-15} 
\end{flushright}
\vspace{1cm}
\begin{center}
{\large \bf Charge cross-over at the $U(1)$-Higgs phase transition}\\[9.5ex]
 
{Filipe Freire} ${}^{a,b}$\footnote{freire@thphys.may.ie}
{and} 
{Daniel F. Litim} ${}^{c}$\footnote{Daniel.Litim@cern.ch\\ 
${}\quad\!$Present address: Theory Division, CERN, CH -- 1211 Geneva 23.}\\[2ex]

${}^a${\it Department of Mathematical Physics, N.U.I. Maynooth, Ireland.\\[2ex]

       ${}^b$School of Theoretical Physics,
       Dublin Institute for Advanced Studies\\
       10 Burlington Road, Dublin 4, Ireland.}\\[2ex]

${}^c${\it  Institut f\"ur Theoretische Physik, Philosophenweg 16\\
  D-69120 Heidelberg, Germany.}\\[6ex]

{\small \bf Abstract}\\[2ex]
\begin{minipage}{14cm}{\small
The type-I region of phase transitions at finite temperature of the
$U(1)$-Higgs theory in 3+1 dimensions is investigated in detail using
a Wilsonian renormalisation group. We consider in particular the
quantitative effects induced through the cross-over of the
scale-dependent Abelian charge from the Gaussian to a non-trivial
Abelian fixed point. As a result, the strength of the first-order phase
transition is weakened. Analytical solutions to approximate flow 
equations are
obtained, and all characteristics of the phase transition are
discussed and compared to the results obtained from perturbation
theory. In addition, we present a detailed quantitative study
regarding the dependence of the physical observables on the
coarse-graining scheme. This results in error-bars for the
regularisation scheme (RS) dependence. We find quantitative evidence
for an intimate link between the RS dependence and truncations of flow
equations.\\[5ex]
}
\end{minipage}
\end{center}
\newpage
\pagestyle{plain}
\setcounter{page}{1}

\renewcommand{\thefootnote}{\arabic{footnote}}
\setcounter{footnote}{0}

\section{Introduction}
\noindent
The phase transition of the $U(1)$-Higgs theory in 3+1 dimensions at
finite temperature provides an important model for cosmological phase
transitions. In the high temperature limit, it reduces to the purely
$3d$ Abelian Higgs model describing the superconducting phase
transition \cite{LuMa}, or certain nematic to smectic-A phase
transitions in liquid crystals \cite{deGennes}. The phase transition
in this model is governed by the infra-red region of its spectrum of
fluctuations. The nature of the phase transition depends primarily on
the ratio $m_{\rm H}/m_{A}$ between the scalar and the gauge field
mass. For superconductors, these mass scales correspond to the inverse
correlation length and the inverse London penetration depth,
respectively.  For small values of the Higgs field mass the phase
transition is strongly enough first order to cut-off long range
fluctuations. This corresponds to the good type-I region for standard
superconductors, $m_{\rm H}/m_{A}<1$. On the other hand, the type-II
region corresponds to $m_{\rm H}/m_{A}>1$. Here, it is expected that
the phase transition changes from first to second order.
\\[1ex]
A proper treatment of the long range fluctuations is decisive  for an
understanding of the $U(1)$-Higgs phase transition as they  change
the effective interactions between the fields. The  `microscopic'
physics in the ultra-violet  is characterised by the couplings at the
short distance length scale $1/\La$ (or $\sim 1/T$, with $T$ the 
temperature). In turn, the
physics close to a first-order phase transition depends typically on
the (small) photon mass $m_A$, and thus requires the knowledge of the
couplings at scales $\ll T$.\footnote{When speaking of `scales' we have
always  `mass scales' or `momentum  scales' $\sim k$ in mind. The
corresponding `length scales' are given  as $\sim k^{-1}$.}
\\[1ex]
A field-theoretical approach which in principle is able to deal with
the effects of long range fluctuations  and which describes the
related scaling of  the couplings is given by the Wilsonian
renormalisation group 
\cite{Wegner73,Polchinski84,Wetterich91,ReuterWetterich93}. This
procedure is based on integrating-out infinitesimal momentum shells
about some `coarse-graining' scale $k$ within a (Euclidean)
path-integral  formulation.  The infra-red effective theory obtains
upon integrating the resulting  flow w.r.t.~$k\to 0$. This way, the
characteristic scaling behaviour  (or `running') of the couplings as
functions of $k$, and in particular  the running of the Abelian charge
$e(k)$, is taken into account.  A Wilsonian approach thus improves  on
perturbative resummations in that the perturbative expansion parameter
$e^2 T/m_A$ becomes now scale dependent,
$e^2(k)T/m_A(k)$. While the former diverges close to a second-order 
phase transition where the photon mass vanishes, the latter remains 
finite in the infra-red even for $m_A(k)\to 0$ 
due to the non-trivial scaling of the Abelian charge.
The crucial role of running couplings in finite temperature phase
transitions has been discussed in pure scalar theories\cite{OSF93,Tetradis93}. 
\\[1ex]
In the present article we employ the Wilsonian renormalisation group
to the type-I regime of the $U(1)$-Higgs phase transition. Our main
contributions are two-fold. First, we take into account the
non-trivial scaling of the Abelian charge $e^2(k)$, characterized by an
effective Abelian fixed point which is kept as a free parameter. The
infra-red  effects lead the Abelian gauge coupling to 
cross-over\footnote{This cross-over is not to be confused with
the qualitatively different `cross-over' observed in the type-II regime
of 3+1 dimensional $SU(2)$+Higgs theory.} from
its slow  logarithmic running in the ultra-violet (effectively $4d$)
to a strong  linear running in the infra-red 
(effectively $3d$).  The
characteristic  scale for this cross-over  depends on the precise
infra-red (IR) behaviour of the Abelian charge, and is  decisive for
both the strength of the transition and the properties  of the phase diagram.
This is currently the least well understood part of the problem.
Equally important is to retain the full field  dependence of the
effective potential (no polynomial approximation), for which an
analytical expression is given in the sharp cut-off case.  We obtain
all thermodynamical quantities related to the first-order phase
transition and study their dependence on the cross-over behaviour.
Second, we present a detailed quantitative analysis of the
`coarse-graining' dependence of our results. This is an important
consistency check for the method and the  approximations involved. We
give quantitative evidence for an intimate  link between a truncation
of the effective action, and the dependence  on the coarse graining
scheme, which can simply be displayed as  additional `error-bars' due
to the scheme dependence.
\\[1ex]
The $3d$ $U(1)$-Higgs phase transition has been studied previously
using flow equations  \cite{BFLLW,BLLW,TL96,Tetradis97,Litim97}, and
within perturbation theory \cite{PT1,Hebecker93,GI}. Recent results
from lattice simulations for both type-I and type-II regions have been
reported as well \cite{Dimopoulos:1998cz,lattice}.  
In \cite{BFLLW}, the RG flow has been studied for
the type-II regime within a local polynomial approximation for the
effective potential about the asymmetric vacuum up to order $\phi^8$
in order to establish the phase diagram, the relevant  fixed points
and the related critical indices. The polynomial  approximation is
expected to give reliable results for the scaling solution close to a
second-order fixed point. The type-I regime has been discussed for the
full potential,  using a matching argument for the running of the
Abelian charge. In \cite{BLLW}, the large\,-$N$ limit and its
extrapolation  down to $N=1$ has been considered as well. It was
pointed out that the local polynomial approximation becomes
questionable close to a first-order phase transition or a tri-critical
fixed point at about $N\approx 4$ or smaller. This was later confirmed
by Tetradis \cite{Tetradis97}, who  in addition abandoned the local
polynomial approximation.  The present study, aiming particularly at
the type-I region of the phase diagram, improves on \cite{BFLLW,BLLW}
in that the full field dependence of the effective potential will be
taken into  account. A quantitative description of thermodynamical
observables at the phase transition requires a good accuracy for the
effective potential in the first place. Our study also goes beyond the
work of \cite{Tetradis97} in three important aspects. We study the 
dependence of physical observables on the value of the
effective Abelian fixed point. In addition, explicit analytical
solutions to approximate flow equations are given, as well as a
discussion of the scheme dependence.
\\[1ex]
This article is organised as follows. We introduce the Wilsonian flow
equations and the particular Ansatz used for the $U(1)$-Higgs theory.
The flows for the Abelian charge and the free energy are explained, as
well as the further approximations involved (Sect.~\ref{flows}). We
then  proceed with the thermal initial conditions as obtained from
perturbative  dimensional reduction (Sect.~\ref{initial}) and 
a discussion of the  phase diagram and the critical line
(Sect.~\ref{phasediagram}). This is  followed by a computation of all
relevant thermodynamical quantities at  the first-order phase
transition as  functions of the effective Abelian fixed point, a
computation of the corresponding characteristic scales, and a
discussion of  the approximations made (Sect.~\ref{PT}).  A
quantitative study of the scheme dependence on the main
characteristics of the phase transition is given
(Sect.~\ref{schemes}), followed by a summary and an outlook
(Sect.~\ref{discussion}). Three Appendices contain some more technical
aspects of our analysis.

\section{Flow equations}\label{flows}

\subsection{Wilsonian flows}
\noindent
Wilsonian flow equations are based on the idea of a
successive integrating-out of momentum modes of quantum fields within
a path-integral formulation of quantum field theory 
\cite{Wegner73,Polchinski84}. This procedure, in turn, can also be
interpreted as the step-by-step averaging of the corresponding fields
over larger
and larger volumes, hence the notion of `coarse-graining'. 
The modern way of implementing a coarse-graining 
within a path-integral formalism goes by adding 
a suitable regulator term $\sim \int\phi R_k(q)\phi$, 
quadratic in the fields, to the action \cite{Wetterich91}.
This additional term introduces a new scale parameter $k$, the 
`coarse graining' scale. A Wilsonian flow equation 
describes how the coarse grained effective action $\Ga_k$ 
changes with the scale parameter $k$, relating this scale dependence
to the second functional derivative of $\Ga_k$ and the scale 
dependence of the IR regulator function $R_k$. 
The boundary conditions for the flow
equation are such that the flow relates the microscopic
action $S=\lim_{k\to \infty}\Ga_k$ with the corresponding
macroscopic effective action $\Ga=\lim_{k\to 0}\Ga_k$,
the generating functional of 1PI Green functions. \\[1ex]
To be more explicit, we follow the `effective average action' approach
as advocated in \cite{Wetterich91} and consider the flow equation 
\beq \label{general}
\0{\partial}{{\partial t}}\Ga_k[\Phi]
=\012\Tr\left\{\left(\Ga^{(2)}_k[\Phi]+
 R_k\right)^{-1}\0{\partial R_k}{{\partial t}}\right\}. 
\eeq
Here,  $\Phi$  denotes bosonic fields and  $t = \ln k$ the logarithmic
scale parameter. The length scale $k^{-1}$ can be interpreted as a
coarse-graining scale \cite{ReuterWetterich93}.  The right hand side
of \Eq{general}  contains the regulator function $R_k$ and the second
functional derivative of the effective action with respect to the fields. The
trace denotes a summation over all indices and integration over all
momenta. The above flow interpolates between the
classical and quantum effective action due to some properties of
the regulator functions $R_k$ (see
Sect.~\ref{regulator}). It is important to realise that the
integrand of the flow equation \eq{general}, as a function of momenta $q$,
is peaked about $q^2\approx k^2$, and suppressed
elsewhere. Consequently, at each infinitesimal 
integration step $k\to k-\Delta k$ only a narrow window of
momentum modes contribute to the change of $\Ga_k\to \Ga_{k-\Delta k}$. 
In particular, modes
with momenta $q\gg k$ do no longer contribute to the running
at the scale $k$. It is this property which justifies the interpretation
of $\Gamma_k$ as a coarse-grained effective action with modes $q\gg k$
already integrated out.
\\[1ex]
For gauge theories, the flow equation
\eq{general} has to be accompanied by a modified Ward
identity which has to be satisfied at each scale $k$.  Such a requirement
is necessary to guarantee that the physical Green
functions obtained for $k\to 0$ obey the usual Ward identity
\cite{Ellwanger,FW,LP1,LP2}. Here, we use the background field
formalism, as employed in \cite{ReuterWetterich93} in the context of
the effective average action with a covariant gauge fixing (Landau
gauge). 
\\[1ex]
The flow equation couples the infinite number of operators describing
an effective action with its second functional derivate. 
In order to solve \eq{general}, one has to truncate $\Ga_k$
to some finite number of operators relevant for the problem under
investigation. Some systematic expansions for the flow equations are known.
Apart from a weak coupling expansion, which is known to reproduce the 
standard perturbative loop expansion, one can use expansions in powers 
of the fields, derivative expansions, or combinations thereof. These
latter expansions have the advantage of not being necessarily restricted
to a small coupling regime. A discussion on the use of a
derivative expansion in Wilsonian RG is presented in \cite{Morris94}.
\\[1ex]
We now turn to our Ansatz for the Abelian Higgs model. The most important
information regarding the phase structure of the model is encoded in the
effective potential (or coarse grained free energy) $U_k$, from which all
further thermodynamical quantities are derived. Equally important is the
wave function renormalisation factor of the gauge fields $Z_F$, which 
encodes the non-trivial running of the Abelian charge. In turn, 
the wave function renormalisation factor $Z_\varphi$ for the scalar fields
is less important because the scalar field anomalous dimension remains
small in the type-I region of phase transitions. Hence, we approximate 
the effective action $\Ga_k$ to leading order(s) in a derivative expansion 
through the following operators
\beq \label{ansatz}
\Ga_k[\phi,A]=\int
d^dx\left\{U_k(\rb)+\014Z_{F,k}F_{\mu\nu}F_{\mu\nu}+
Z_{\varphi,k}(D_\mu[A]\varphi)^*D_\mu[A]\varphi \right\}
\eeq
where $\rb=\varphi^*\varphi$, $F_{\mu\nu}=
\partial_\mu A_\nu - \partial_\nu A_\mu$ is the field strength of the
electromagnetic field, and ${D}_\mu$ denotes the covariant
derivative $\partial_\mu-i\bar e A_\mu$. 
\\[1ex]
In principle, 
the flow equation can be used directly (starting with initial
parameters of the 4$d$ theory at $T=0$) to compute the corresponding 
critical potential at finite temperature within the imaginary time 
formalism, or, like in \cite{Litim98}, using a 
real-time formulation of the Wilsonian RG \cite{Pietroni}. 
Our strategy in the present case is slightly simpler. 
We are interested in the region of 
parameter space where the $4d$ couplings are small enough to 
allow a perturbative integrating-out of the super-heavy and heavy 
modes, i.e.\,the non-zero Matsubara modes for all the fields and 
the Debye mode. In this case, we can rely on the dimensional reduction
scenario and employ the results of  
\cite{Kajantie96a}, who computed the initial conditions  perturbatively.
The result is then a purely 
three-dimensional theory for the remaining light degrees of 
freedom, whose infra-red behaviour is then studied applying the 
above Wilsonian renormalisation group. 
In the sequel, we will therefore need the flow equations for $U_k$ and 
$e^2(k)$ in $3d$. At the scale of dimensional reduction, that is the 
starting ultra-violet (UV) scale $\La$ of the $3d$-flow, we normalize 
the wave function factors to one, and the initial effective potential 
$U_\La$ is obtained from dimensional reduction. 

\subsection{Cross-over of the gauge coupling}\label{runninge}
\noindent
We now consider the case $d=3$, and discuss the flow for the Abelian 
coupling. A main feature of the Abelian Higgs theory in $3d$ is that the
Abelian charge scales in a non-trivial manner with the coarse graining
scale $k$. The dimensionless Abelian charge in $3d$ is defined as
\beq
\e2_3(k)=\0{\eb2_3(\La)}{Z_F(k)k}\equiv \0{\eb2_3(k)}{k}\ ,\eeq 
and its scale dependence is related to the gauge
field anomalous dimension $\eta_F=-\partial_t \ln Z_F(k)$ 
(here a function of $k$ and the fields) through 
\cite{ReuterWetterich93}  
\beq \label{flowe2} \0{d
\e2_3}{dt}=-\e2_3\,(1 - \eta^{}_F)\ .  
\eeq 
The first term in \eq{flowe2} comes from the intrinsic dimension of the 
charge squared (proportional to $k$), 
while the second term proportional to the gauge field 
anomalous dimension accounts for the non-trivial running of the coupling. 
The flow \eq{flowe2} has always the (trivial) Gaussian fixed point 
given by $\e2_3=0$. In addition, one might encounter further non-trivial 
fixed points which are given implicitly through the solutions of
$\eta_F=1$. 
\\[1ex]
Both the scalar and the gauge field anomalous dimensions 
$\eta_\F$ and $\eta_F$ are perturbatively small near the 
Gaussian fixed point, i.e.~$|\eta_\F|$ and $|\eta^{}_F| \ll 1$. This
holds true at the initial scale for $k=\La$ in the effective 
{\it 3d\/} running to be specified later.
It follows that the running of the dimensionful Abelian
charge is negligible near the Gaussian fixed point, 
$\eb2(k)\approx\eb2(\Lambda)$. Here, the dimensionless coupling scales as
$e^2_3(k)\sim \La/k$. In this regime it is expected that standard
perturbation theory gives a reliable estimate of the effective 
potential in this region of the parameter space \cite{Litim97}. 
\\[1ex]
However, for $\eta_F<1$ the Gaussian
fixed point is IR unstable, as follows directly from \eq{flowe2}. Therefore,
when approaching the infra-red, the dimensionless Abelian charge will 
unavoidably grow large, scaling away from the Gaussian fixed point. In 
particular, it can enter into a region where $\eta_F(e^2)$ is no 
longer $\ll 1$.  When a non-trivial fixed point is approached, 
i.e.~$\eta_F\approx 1$, the suppression factor 
$(1-\eta_F)$ in \eq{flowe2} becomes important. A 
strong linear running of  $\eb2\sim k$ (the IR  region is effectively $3d$)  
will ultimately set in as soon as the deviation from the Gaussian
fixed point becomes sizeable \cite{BFLLW,BLLW,Litim94a}. In this regime,
we expect some quantitative modifications of the predictions by 
perturbation theory due to the non-trivial running of the Abelian charge.

\subsection{Abelian fixed point}\label{effective}
\noindent
The anomalous dimension $\eta_F$ has been calculated in
\cite{ReuterWetterich93}. It is, in general, a complicated function of
the gauge coupling, the fields, and the further parameter describing the 
effective action in a given approximation, like the coarse grained 
potential (cf.~(113) of \cite{BFLLW}). However, $\eta_F$ is proportional 
to $\e2_3$ itself, and we write it as 
\beq
\label{etaF} \eta^{}_F(\bar \rho)= \frac{\,\e2_3\,}{\es2(\bar \rho)} \,.
\eeq
Given the anomalous dimension, \eq{etaF} provides a definition of 
$\es2(\bar \rho)$. Our current understanding of the IR behaviour 
of the gauge sector hinges on the precise properties of 
$\eta_F(\bar \rho)$,  and hence of $\es2(\bar \rho)$. 
\\[1ex]
Let us recall a few cases where $\es2$ is approximately known. 
First, within standard perturbation theory, the dimensionful gauge 
coupling ${\bar\e2_3}= {\e2_3}{k} = const$ throughout. Within our 
formalism, the `no-running' corresponds to the limit $\es2\to\infty$. 
In this limit, the effective fixed point is independent of the fields 
and we can expect to be close to the results from  perturbation theory, 
as long as
additional effects due to the scalar anomalous dimension can be 
neglected.\footnote{In the region where $\lambda \gg e^2$
  (e.g. strongly type-II superconductors), the critical behaviour of 
the limit $\es2\to\infty$ corresponds to an effective scalar theory which
belongs to a different universality class than the $O(2N)$ scalar theory 
obtained for $\es2\to 0$ \cite{BFLLW}.} 
\\[1ex]  
Second, consider the large\,-$N$ limit of the $U(1)$-Higgs model, 
where $N$ denotes the number of complex scalar fields. In this limit, 
the flow \eq{flowe2} is dominated by the contributions of the 
Goldstone modes. They overwhelm those due to the radial mode. 
Therefore, $\es2$ becomes
\beq
\label{es2N} \es2 \approx  \frac{6 \pi^2}{N} 
\eeq
close to the minimum of the effective potential.\footnote{In 
\eq{es2N}, `$\approx$' means equality up to a regulator scheme 
dependent coefficient of ${\cal O}(1)$.}  In particular, 
\eq{es2N} does no longer depend on the quartic scalar coupling 
or the location of the v.e.v.~because the massive (radial) mode 
is suppressed. Extrapolating \eq{es2N} down to the physically 
relevant case $N=1$ corresponds to replacing the radial mode 
by a massless one. This yields $\es2 \approx 6 \pi^2$ in 
accordance with the leading order result from the 
$\epsilon$-expansion. This value serves as a reference value for 
our subsequent considerations. 
\\[1ex]
Third, we recall the findings of \cite{BFLLW} and \cite{BLLW}, 
where the function $\eta_F$ has been studied numerically for 
different $N$ within a local polynomial approximation of the 
flow about the non-trivial minimum at $\bar\rho=\bar\rho_0$  
(up to $\sim \phi^8$). It was found that the implicit solutions 
to $\eta_F(\es2)=1$ for small $N$ (in particular $N=1$) can 
deviate considerably from the large\,-$N$ extrapolation $6\pi^2$. 
This deviation is due to the decoupling effects of the massive 
mode. Still, the qualitative form of \eq{etaF}, where the function 
$\es2$ is replaced by an effective field-independent fixed point, 
remains a good approximation to \eq{etaF}. This simplified picture 
persists if the field derivatives 
$\partial\ln[\es2(\bar\rho)]/\partial\bar\rho$ remain small within 
the non-convex region of the effective potential (see also the 
discussion in Sect.~\ref{error}). This implies that the threshold 
effects of the radial mode for $N=1$ act on \eq{flowe2} as varying 
the number of scalar fields in \eq{es2N}.
\\[1ex]
Hence, the qualitative structure of the flow \eq{flowe2}, to leading 
order, is determined by \eq{etaF} with $\es2$ given by some number, 
{\it e.g.}~the appropriate effective fixed point. For the present 
purpose it is sufficient to study the flow \eq{flowe2} with $\es2$ 
as a free parameter. The properties of the first-order phase transition 
depend on the size of $\es2$. However, as we shall see in detail below, 
the dependence turns out to be very small for large $\es2$: this part of 
the phase diagram can be studied without having a complete understanding 
of the underlying fixed point structure. In turn, we find a strong 
dependence within regions where the effective fixed point is small. 
For this case, a more refined analysis is required in order to 
provide more reliable predictions.  

\subsection{Cross-over scale}\label{cross-over}
\noindent
Within the remaining part of the article we approximate the anomalous 
dimension as described above. Hence, the eqs.~\eq{flowe2} 
and \eq{etaF} are easily solved by
\beq\label{e2k}
\e2_3(k)=\0{\,\es2}{1 + k/\kcr}\ .
\eeq
We note the appearance of a characteristic {\it cross-over scale} 
\beq\label{ktr1}
\kcr= \0{\La\, e^2_3(\La)}{\,\es2-e^2_3(\La)}\ .
\eeq
It describes the  cross-over between the Gaussian and the Abelian
fixed point, and depends on the initial conditions. For
$k> \kcr$ the running is very slow and dominated by the Gaussian fixed
point, $\eb2_3(k)\sim const$\,. This corresponds also to the 
limit $\es2\to \infty$. On the other hand, for $k<\kcr$ the running
becomes strongly linear and the Abelian fixed point governs the
scale dependence, $\eb2_3(k)\sim k$. The question as to how strong
the first-order phase transition is affected by this cross-over
depends on whether the cross-over scale is much larger (strong effect)
or much smaller (weak effect) than the typical scales of the transition
(see Sect.~\ref{char-scales}).
\begin{figure}[t]
\begin{center}
\unitlength0.01\hsize
\begin{picture}(50,15)
\epsfxsize=0.5\hsize
\epsffile{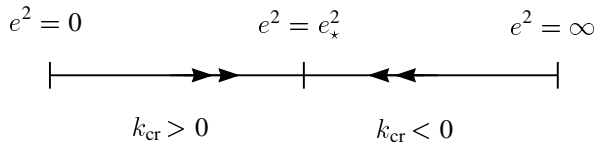}
\end{picture}
\begin{minipage}{.9\hsize}{\caption{\label{e2fp}
      \small The relation 
between the sign of the cross-over scale $\k-{cr}$ and the running of 
the gauge coupling
(arrows indicate the direction of the $e^2$ flow as $k\to 0$).}}
\end{minipage}
\end{center}
\end{figure}
\noindent
The cross-over scale turns negative if the initial value $e^2_3(\La)$
is too big. This simply means that the flow would never be dominated by the
Gaussian fixed point (see Fig.~\ref{e2fp}) in the first place 
(no cross-over). Although this case is interesting 
in its own right, this region will not be discussed any further.

\subsection{The running potential}\label{runningp}
\noindent
We now turn to the flow equation for the effective potential, which
can be obtained from the flow equation \eq{general} using the Ansatz
given by \eq{ansatz}.  The resulting flow equation is a second-order
non-linear partial differential 
equation. It has been derived originally in \cite{ReuterWetterich93} 
and reads in 3$d$
\beq\label{flowpot}
\frac{4 \pi^2}{k^2}\,\0{d\,}{dk}\,U_k(\rb)=(2N-1)\,\el 30
\left(\0{U'_k(\rb)}{k^2}\right) +\el 30
\left(\0{U'_k(\rb)+2\rb U''_k(\rb)}{k^2}\right) + 
2\,\el 30 \left(\0{2\eb2_3(k)\rb}{k^2}\right)\ 
\eeq
for the case of $N$ complex scalar fields. Similar flow equations are obtained
for the wave function factors $Z_\varphi$ and $Z_F$, and thus for
the anomalous dimensions $\eta_\varphi=-\partial_t \ln Z_\varphi$ and
$\eta^{}_F=-\partial_t \ln Z^{}_F$. Here, $\el 30(\om)$ denotes a
scheme dependent threshold function defined as
\beq\label{ldn1}
\ell^d_n(\om)=
-\left(\de_{n,0}+n\right)\int^\infty_0 dy 
 \0{r'(y)y^{1+\s0d2}}{[y(1+r)+\om]^{n+1}} \ .
\eeq
These functions have a pole at some $\omega<0$ and vanish for large arguments.
The function $r(q^2/k^2)$ is related to the regulator function $R_k$ 
introduced in \eq{general} through
\beq\label{reg}
R_k(q^2)= Z\ q^2\ r(q^2/k^2),
\eeq
where $Z$ denotes either the scalar or gauge field wave function
renormalisation.\footnote{A more detailed discussion of both $R_k$ and the
dimensionless functions $r(q^2/k^2)$ is postponed until
Sect.~\ref{regulator}.} 
\\[1ex]
We can distinguish three different contributions to the running of the
potential \eq{flowpot} which are, from the left to the right, related to the
massless scalar, massive scalar, and the gauge field fluctuations,
respectively. Not all the three of them are of the same order of
magnitude, though. Indeed, it was already noted
\cite{ColemanWeinberg73} that the gauge field fluctuations
dominate \eq{flowpot} if the quartic scalar coupling $\la$ is much 
smaller than the gauge coupling squared, $\la/\e2\ll 1$. This is
the case for the physically relevant initial conditions, that is, for the
starting point of the flow equation \eq{flowpot}.
Therefore, we can make a further approximation and
neglect the contributions from the scalar field fluctuations compared
to those from the gauge field.  The flow equation thus takes the form
\beq\label{flowpot1}
\frac{2 \pi^2}{k^2}\,\0{d\,}{dk}\,U_k(\rb) = \el
30\left(\0{2\eb2_3(k)\rb}{k^2}\right)\ .
\eeq
Integrating the 
approximated flow equation allows to control self-consistently whether
the effects from scalar fluctuations remain negligible or not. It
suffices to evaluate the right-hand side of \eq{flowpot} with $U_k$ from the
solution of \eq{flowpot1} to compare the contribution of the neglected
terms to the running of, say, $U''_k$ with the leading
contributions. It is 
well known that the scalar fluctuations are important for the inner
part of the effective potential which becomes convex in the limit
$k\to 0$ \cite{Tetradis92}. Therefore it is to be expected that this
approximation becomes unreliable, within the non-convex part of the
potential, at some scale $k_{\rm flat}$. 
\\[1ex]
The solution to \eq{flowpot1} is  the first step of a
systematic iteration to compute the solution to \eq{flowpot}. The next
step would be 
to replace $U_k$ on the r.h.s.~of \eq{flowpot} by the solution to
\eq{flowpot1}. Proceeding to the next iteration step the scalar
fluctuations are eventually taken into account. Solving \eq{flowpot}
with $U_k$ on the right-hand side replaced by the 
explicit solution of \eq{flowpot1} is much easier than solving
\eq{flowpot} directly, because the former becomes an ordinary
differential equation, while the later is a partial one.  This
procedure can be interpreted as an expansion in terms of scalar loops
around the gauge field sector. We will mainly use the first step in
the sequel. In order to estimate the integrated contribution of the 
scalar fluctuations,
we will in addition discuss the solution of \eq{flowpot} with $U_k$ on
the right-hand side replaced by $U_\La$ (see Appendix \ref{AppC}).

\subsection{The coarse-grained free energy}\label{fe}
\noindent
The coarse grained free energy obtains as the solution to the coupled 
set of flow equations \eq{flowe2} and \eq{flowpot}. In the present case,
a solution can be written as
\begin{mathletters}\label{solution}
\beq\label{Uformal}
U_{k}(\rb) = U_\La(\rb) + \De_{k}(\rb)\ .
\eeq
Here, the term $\De(\rb)$ stems from integrating out the $3d$ 
fluctuations between the scales $\La$ and $k$. With $e^2_3(k)$ 
from \eq{e2k} and $dU_k/dk$ from \eq{flowpot1}, it reads 
\beq\label{Deltadef}
\De_{k}(\rb)=\01{2\pi^2}\int^\La_kd\bar k
\int^\infty_0dy\ \0{r'(y)\,y^{5/2}\,\bar k^3(1+\bar k/\kcr)}{y\ \bar k 
(1+r)\,(1+\bar k/\kcr)+2\es2\rb\, } +{\rm const.}  
\eeq
\end{mathletters}%
The constant is fixed by requiring that $\De_{k}(0)=0$. In Eq.~\eq{Deltadef} 
we see that the term resulting form integrating-out $3d$ effective
modes depends on the RS through the regulator function $r(y)$ and its
first derivative. (Explicit expressions are given in the Appendix \ref{AppB}).
\\[1ex]
The above expressions are enough to study all properties of the phase 
transitions as functions of the parameters of the potential $U_\La$. 
\\[1ex] 
We are aiming to use an initial condition at $k=\La$ obtained 
from perturbation theory in $4d$.  This requires that the
parametrisation of the $3d$ potential $U_\La$ is such that the
matching equates the right parameters.
In the universal limit $\La\to\infty$, the effective mass term
contained via $U_\La$ is renormalised  to $U_\La\to U_\La-C_\La\rb$. 
For a sharp cut-off, we find explicitly 
\beq\label{C}
C_\La(e)=
        \0{e_\star^2}{\pi^2}
        \left(\,\Lambda \k-{cr} -\k-{cr}^2 \ln(\Lambda/\Lambda_0)\,\right)\ .
\eeq
For finite $\La$, this corresponds to a finite renormalisation of 
the parameters of the theory, {\it i.e.} the mass term, or, equivalently,
a finite shift of the v.e.v.~at the matching scale.\footnote{This shift 
corresponds to the finite renormalisation as employed 
in \cite{Tetradis97}.} This finite
renormalisation has its origin simply in the way how the flow equation
integrates-out the $3d$ momentum scales. Only after this
transformation it 
will be appropriate to identify the potential $U_\La$ at the scale of 
dimensional reduction with the renormalised effective potential 
obtained from a perturbative calculation. 

\section{Thermal initial conditions}\label{initial}
\noindent
We now specify in concrete terms the initial conditions for the effective
$3d$ theory. 
The task is to relate the $3d$ renormalised parameters of the 
effective potential to those of the $T=0$ $4d$ theory. 
The initial conditions for the $3d$ running 
are the potential $U_\La(\rb)$ and the gauge coupling $\eb2_3(\La)$.  
The effective perturbative $3d$  Lagrangean has been derived 
in \cite{Kajantie96a}.  We start with the $4d$ effective action,
\beq\label{lag}
\Ga[\phi,A]=\int d^4x\left\{\0{1}{4}\,F_{\mu\nu}F_{\mu\nu}+({\cal
D}_\mu\phi)^\dagger 
({\cal D}_\mu\phi) - \frac{m_{\rm H}^2}{2}\,\phi^\dagger\phi +
\frac{\la}{2}\,(\phi^\dagger\phi)^2\right\} ,
\eeq
where $\phi$ is a single component $4d$ complex scalar field. 
The mass parameter $m_{\rm H}$ entering \eq{lag} denotes the 
$T=0$ Higgs boson mass. It is related to the other zero temperature 
parameters of the theory by
\beq\label{t0vev}
\0{\la}{\e2} = \0{m_{\rm H}^2}{M^2_{\rm W}}
\eeq
with $M_{\rm W}$ the photon mass. In the phase with 
spontaneous symmetry breaking, $m^2_{\rm H}>0$, we have 
$\langle\,\phi^*\phi\,\rangle\,\equiv\,v^2/2 =  M^2_{\rm W}/2e^2$.  
The effective action for the 3$d$ theory obtains as 
\begin{mathletters}\label{ULa}
\bea
\Ga_\La[\F,A]&=&\int d^3x\left\{\0{1}{4}\,F_{ij}F_{ij}+({\cal
D}_i\F)^\dagger({\cal D}_i\F) 
+V_\La(\rb) \right\}\ ,\label{GaLa}\\[2ex]
V_\La(\rb)&=&m_3^2\,\F^\dagger\F + \frac{\bar\la_3}{2}\,(\F^\dagger\F)^2
\eea
\end{mathletters}%
where $\F$ is the static component of $\phi$ and $i,j$ the spatial 
components of $\mu,\nu$. The electric component of the gauge field has been 
fully integrated out because it acquires a thermal (Debye) mass
$m_D$. The effects of the fluctuation of this 
mode are suppressed by inverse powers of $T$ as
$m_{\rm D} \propto T$, like the nonstatic modes. 
Following \cite{Kajantie96a}, the matching conditions read to 1-loop accuracy
\begin{mathletters}\label{matching}
\bea
\bar\e2_3(\La)& =& \e2 T\,\,
\label{match1}\\
\bar\la_3(\La) &=& \left(\la + \0{\e4}{4\pi^2}\right) T - 
\0{\e4}{4\pi}\,\,\frac{T^2}{m_{\rm D} (\La)} \,\,
\label{match2}\\
m^2_3(\La) &=& \left(\014\,\e2+\016\la\right)T^2 - \01{2}\, m_{\rm H}^2 -
\0{\e2 }{4\pi}\,T\,m_{\rm D}(\La)\,\,
\label{match3}\\
m^2_{\rm D}(\La) &=& \013\, \e2 T^2\ .\,\,
\label{Debye} 
\eea
\end{mathletters}%
Using the above, and taking into account the finite 
renormalisation \eq{C} as explained in Sect.~\ref{fe}, the renormalised 
effective initial potential $U_{\La}(\rb)$ entering \eq{Uformal} 
can be expressed in terms of the $T=0$ 
parameters and \eq{matching} as
\begin{mathletters}\label{initialpot}
\beq
U_{\La}(\rb) =
-\,m^2_{\rm R}\,\rb+\012\, \bar\la_{\rm R}\,\rb^2
\eeq
with
\bea \label{A}
m^2_{\rm R}(\La)&=&
\012\,m^2_{\rm H}-\left(
\0{\e2}4+\0{\la}6 - \0{\e3}{4\sqrt{3}\pi}\right)T^2
                              + C_\La(e)\ ,
\\   \label{B}
\bar\la_{\rm R}(\La)&=&\left( 
\la + \0{\e4}{4\pi^2} - \0{\sqrt{3}\,\e3}{4\pi}\right)T \ ,
\eea
and the dimensionless renormalised quartic coupling reads 
$\la_{\rm R}=\bar\la_{\rm R}/\La$. The renormalised v.e.v.~at the 
scale of dimensional reduction follows as
\beq\label{rhoR}
\rb_{\rm R}(\La)={m^2_{\rm R}}(\La)/{\bar\la_{\rm R}}(\La)\ .
\eeq
\end{mathletters}%
All the $3d$ parameters are now defined at the reduction scale 
$\La$, which is on dimensional grounds linearly related to 
the temperature,
\beq\label{LaT}
\La = \xi\ T \ .
\eeq  
Using \Es{ktr1}, \eq{match1} and \eq{LaT} it follows, that the
cross-over scale $\kcr$ is also related to $T$ as
\beq\label{ktr2}
\kcr = \0{\xi\, e^2}{\,\xi\,\es2 -e^2}\ T\ .
\eeq
Let us finally comment on the matching parameter $\xi$.
On one hand, $\xi$ has to be smaller 
than $2\pi$, because elsewise the assumption that all heavy modes
 have been integrated out can no longer be maintained. On the other 
hand, a too small value for $\xi$, say $\xi < 1$, would tend to neglect 
contributions from modes roughly within the window $\approx 2\pi T$ 
and $\approx T$. For the problem under consideration $\xi\approx 1$ 
turns out to be a good choice. This choice shall be adopted throughout.
Our results do depend very little on a variation of this matching scale
(see also the comment in Sect.~\ref{char-scales} below).

\section{The phase diagram at finite temperature}\label{phasediagram}
\noindent
We have now all the ingredients to study in detail the phase diagram and the
phase transition of scalar electrodynamics. In this section, we discuss
the main characteristics of the phase diagram as well as some properties 
of the critical line. The following section collects our results for the 
thermodynamical quantities related to the first-order phase transition
and a discussion of the characteristic scales of the problem.
\begin{figure}
\begin{center}
\unitlength0.001\hsize
\begin{picture}(700,600)
\put(65,530){{\huge $\0{T}{M}$}}
\put(300,75){{\Large $m_{\rm H}$}[GeV]}
\put(200,450){\Large SYM}
\put(400,200){\Large SSB}
\put(190,200){\large 1${}^{st}$}
\put(440,490){\large 2${}^{nd}$}
\hskip100\unitlength
\psfig{file=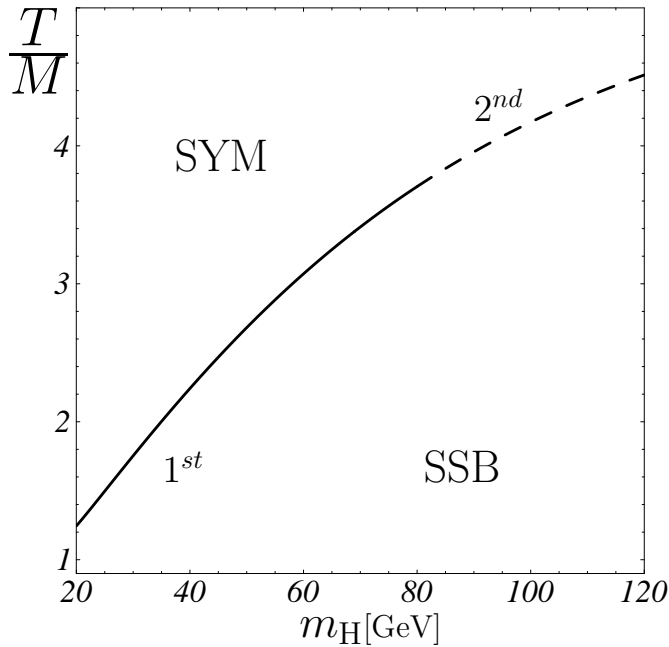,width=500\unitlength}
\end{picture}
\vskip-.5cm
\begin{minipage}{.8\hsize}
\caption{\label{pd} \small  The phase diagram in the $(T,m_{\rm H})$-plane.}
\end{minipage} 
\end{center}
\end{figure}
\noindent

\subsection{The phase diagram}
\noindent
The `phases' of scalar electrodynamics are distinguished by the location
of the global minimum of the effective potential. Above the critical 
temperature, the ground state 
corresponds to vanishing field $\rb_0= 0$, that is, to the symmetric 
phase (SYM).
Below the critical temperature, the ground state corresponds to $\rb_0\neq 0$,
the phase with spontaneous symmetry breaking (SSB).\footnote{It is
sensible to speak of two distinct phases only for $N>1$ complex scalar 
fields. For $N=1$, the symmetry is never broken in the strict sense.
However, we will stick to the usual\,\,-- albeit slightly incorrect
--\,\,terminology even for $N=1$.}  
The corresponding phase diagram in the $(T,m_{\rm H})$-plane is displayed 
in Fig.~\ref{pd}.
The phase transition between these two phases is first order for small
$\bar \la_3/\eb2_3$, that is for small values of the Higgs field mass. 
In the context of superconductivity this region corresponds to the
strongly type-I systems. 
For very large Higgs field mass, the phase transition
turns second or higher order \cite{BFLLW}.\footnote{The strongly 
type-II region has been studied using flow equations within a local 
polynomial approximation in \cite{BFLLW}. See also \cite{Tetradis97}.} 
\begin{figure}[t]
\begin{center}
\unitlength0.001\hsize
\begin{picture}(1000,500)
\put(460,80){ \Large $\rb/T$}
\put(220,450){\fbox{\Large  $10^7 U_{k,T}/T^3$}}
\put(280,340){$k\approx k_{\rm stable}$}
\put(600,470){$T=T_s$}
\put(585,200){$T=T_b$}
\put(750,320){$T=T_c$}
\put(800,200){$T<T_c$}
\put(700,420){$T>T_c$}
\hskip.04\hsize
\psfig{file=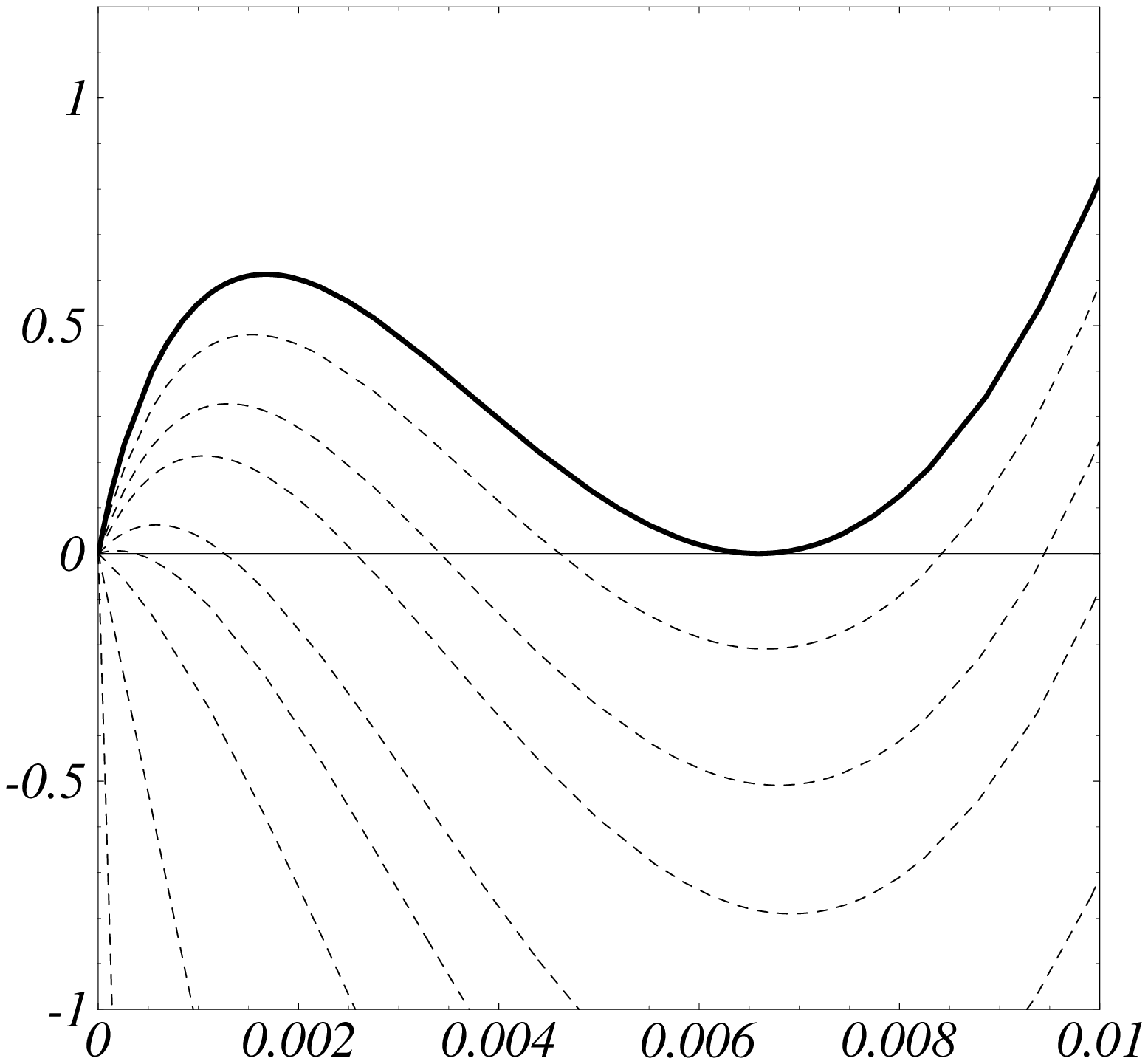,width=.45\hsize}
\psfig{file=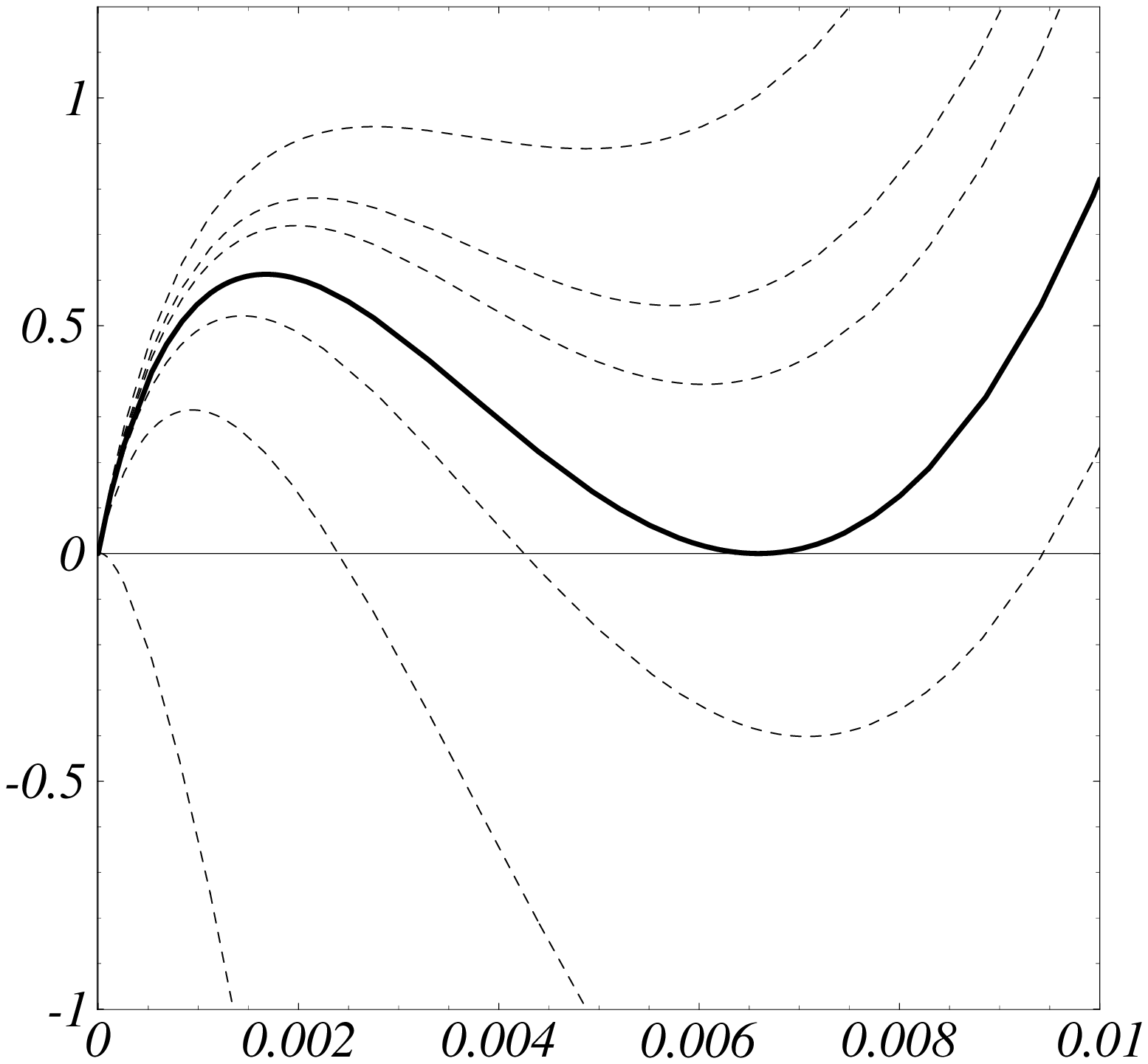,width=.45\hsize}
\end{picture}
\vskip-1.cm 
\begin{minipage}{\hsize}
\caption{\label{Pot-kT} \small The coarse grained free energy 
as a function of the scale parameter and the temperature 
($m_{\rm H}=60\,\,{\rm GeV}$). The full line corresponds to $T=T_c$ 
and $k\approx k_{\rm stable}$. Left panel: $T=T_c$, for different 
scales $k$. Right panel: $k\to k_{\rm stable}$, for different 
temperatures around $T_c$.}
\end{minipage} 
\end{center}
\end{figure}
\noindent
In Fig.~\ref{Pot-kT}, we have displayed the coarse grained free energy 
within the type-I region of parameters for $m_{\rm H}=60$ GeV for different 
scales and temperatures. At the critical temperature (left panel), 
it is realised that a barrier is building up for decreasing 
scale $k$, eventually creating a second minima at vanishing field. 
The minima are degenerate in the infra-red limit 
$k\approx k_{\rm stable}$ (which corresponds roughly to $k\to 0$ 
in the
present approximation). Notice 
that the flattening of the inner part of the potential is not 
observed because the scalar fluctuations have been neglected at 
the present state. Rather, the effective potential reaches the 
degenerate shape already at some scale $k_{\rm stable}$, which should be 
larger than the scale where the flattening 
sets in.\footnote{A quantitative discussion of these scales is 
given in Sect.~\ref{char-scales} below.}
\\[1ex]
The temperature dependence of the coarse grained free energy at  
$k\approx k_{\rm stable}$ is shown in the right panel. The metastability
range $\Delta T=T_s-T_b$ between the barrier temperature $T_b$, where 
the potential 
develops a second minimum at the origin (lowest dashed curve) and the 
spinodal temperature $T_s$, where the asymmetric minimum 
disappears (upper-most dashed curve), is very small.
\\[1ex]
The physical quantities that characterise a first-order phase 
transition (except the metastability range) are defined at the 
critical temperature $T_c$, when the potential has two degenerate 
minima, the trivial one at $\rb=0$ and a non-trivial one at 
$\rb=\rb_0\neq 0$. 
The critical line of the phase diagram as depicted in Fig.~\ref{pd} 
is obtained solving the criticality conditions 
\begin{mathletters}\label{crit0}
\beq\label{minimum}
0= \left.\0{d U_k}{d\rb}\right|_{\rb=\rb_0}
\eeq
\beq\label{deg}
U_k (0)=U_k (\rb_0)\ .
\eeq
\end{mathletters}%
Here we kept $k$ arbitrary though strictly only for $k=0$ are 
these conditions required physically. They establish a relationship 
between the parameters of the theory, and thereby define the critical 
line between the symmetric and the SSB phase in Fig.~\ref{pd}. 
It is helpful to rewrite the conditions \eq{crit0} into 
\begin{mathletters}\label{crit}
\bea\label{crit1}
F_1\left(\rb/T\right)&=&\la_{\rm R}\\
\label{crit2}
F_2\left(\rb/T\right)&=&2\,\0{m^2_{\rm R}}{T^2} \ .
\eea
\end{mathletters}%
The functions $F_1$ and $F_2$ are related to the fluctuation integral 
through
\begin{mathletters}\label{F}
\bea
F_1(x)&=& \02{x^2}\left[\t{\De}(x)-x\t{\De}'(x)\right]\\ 
F_2(x)&=& \0{\,2\,}x\left[2\t{\De}(x)-x\t{\De}'(x)\right] \ ,
\eea
with
\beq\label{Deltatilde}
\t{\De}(\rb/T) =\De(\rb)/T^3 \ .
\eeq
\end{mathletters}%
The first condition determines the ratio  $x=\rb/T_c$ of the 
discontinuity to critical temperature in dependence on the $4d$ 
parameters as given through $\la_{\rm R}(e,\la)$ from \eq{B}. The second 
one relates the solution of \eq{crit1} to the ratio of the Higgs 
boson mass to critical temperature and \eq{A}, and eventually to 
the critical temperature and the discontinuity itself.  
\\[1ex]
Explicit expressions for the scale-dependent effective potential and
the function $\De(\rb)$ are given in the Appendix \ref{AppB}. 

\subsection{Endpoint of the critical line}
\noindent
Some simple properties of the solutions to \eq{crit} can be deduced
directly from the functions $F_{1,2}$. 
For $x>0$, these functions [with $\De_k$ from
\eq{SharpExpl}] are  positive, finite, monotonically decreasing and 
vanishing for $x\to\infty$. They reach their respective maxima at
$x=0$, with (for $k=0$) 
\begin{mathletters}\label{F120}
\bea
F_1(0)&=&\0{2}{\pi^2}\es2 e^2\ ,\\
F_2(0)&=&\02{\pi^2}\0{\xi^2\es2 e^2}{\xi\es2-e^2}
         \left(1-\0{e^2}{\xi\es2-e^2}\ln
           \Bigg(\xi\0{\es2}{e^2}\Bigg)\,\right) .
\eea
\end{mathletters}%
The renormalised $3d$ quartic coupling $\la_{\rm R}$, as given by  
\eq{B} and fixed through the $4d$ parameters of 
the theory, is positive in the domain under consideration. Given 
the monotony property of $F_1$ it follows that a solution to 
\eq{crit1} is unique (if it exists). There exists no solution for 
too large values of $\la_{\rm R}$. Its largest possible value
corresponds to vanishing v.e.v.~{\it i.e.}~to $x=0$. 
Using \Es{t0vev} and \eq{B} gives an upper bound on the scalar mass
for the phase transition being first order. It reads
\beq \label{mh-crit}
\0{m_{\rm H}^2}{M_{\rm W}^2}\le \0{2\es2}{\pi^2} \ .
\eeq
For any finite value of $\es2$ \eq{mh-crit} predicts an upper limit
for the mass of the Higgs particle. This is an immediate consequence
of the existence of an effective fixed point for the running gauge 
coupling \eq{flowe2}. Indeed, as the limit $\es2\to\infty$ corresponds 
to perturbation theory we recover the standard perturbative
prediction of a first-order phase transition for all Higgs boson 
mass. This endpoint is usually interpreted as the tri-critical 
point of the model, above which the phase transition turns from a 
first-order transition to a second-order one. However, the endpoint of 
the first-order transition line is within the
domain of validity of the present computation only for sufficiently 
small values of $\es2$.\footnote{The endpoint presented in Fig.~\ref{pd} 
corresponds to $\es2\approx 5$.} For larger values of the Abelian fixed 
point, we expect that the precise location of the endpoint is also
determined by the scalar field fluctuations.
\\[1ex]
In the opposite case, the smallest possible value for $\la_{\rm R}$
corresponds to $x\to \infty$, thus to $\la_{\rm R}=0$. This gives  a lower
bound on the mass of the Higgs particle according to
\beq \label{higgslow}
\0{\sqrt{3}e}{4\pi}<\0{m_{\rm H}^2}{M^2_{\rm W}}\ .
\eeq
For $M_W=80.6$ GeV and $e=0.3$ the bound is satisfied at 
about $m_{\rm H}\approx 16$ GeV.
This bound stems entirely from the initial conditions employed. This
indicates that the dimensional reduction scenario is no longer appropriate
for small $m_{\rm H}$. In the present work, we are also not interested in
the region of parameter space where the Coleman-Weinberg mechanism 
already takes place within the original $4d$ theory, which happens 
at even smaller values for $m_{\rm H}$ (typically for $\la/e^4$ at
about $3/8\pi^2$ or smaller \cite{Litim94a}).

\section{Thermodynamics of the first-order phase transition}\label{PT}
\noindent
Here we present our results for the coarse grained free  energy and related  
physical quantities close to the critical temperature of the
first-order phase transition as a function of the effective 
Abelian fixed point. The initial
conditions are specified through the gauge coupling at vanishing
temperature $e=0.3$ and the photon mass $M_{\rm W}=80.6$\,GeV. The ratio
$\la/e^2$ of the $4d$ couplings ranges between $0.06-0.75$ for a 
Higgs field mass between $20-70$\,GeV.  
Relevant information is
given by the critical temperature $T_c$, the discontinuity at the phase
transition $\rb_0$, the latent heat $L$ and  the surface 
tension $\si$.\footnote{A comment concerning the
dimensions is in order: $U, \si, L$ and $\rb$ will be given in $3d$
units, unless otherwise stated. Their $4d$ counterparts are simply
obtained by multiplying with $T$.} We compare our findings to perturbation 
theory, and to lattice simulations (for the critical temperature). All 
our results are obtained
as functions of the effective fixed point of the Abelian charge. Due 
to the approximations performed, they depend also on the regularisation 
scheme. We use a sharp cut-off regulator throughout the present section.   
The regularisation scheme dependence is discussed in the following
section.

\subsection{Discontinuity and critical temperature}
\begin{figure}[t]
\begin{center}
\unitlength0.001\hsize
\begin{picture}(700,600)
\put(290,530){ { $4d\!:\ T=0$}}
\put(290,470){ { $3d\!:\ k=\La$}}
\put(290,300){ { $3d\!:\ k\approx k_{\rm vev}$}}
\put(270,85){ {\Large $m_{\rm H}$}[GeV]}
\put(150,200){ \fbox{{\large $\log_{10}(T\rb_0/\rb_{4d})$}}}
\hskip.04\hsize
\psfig{file=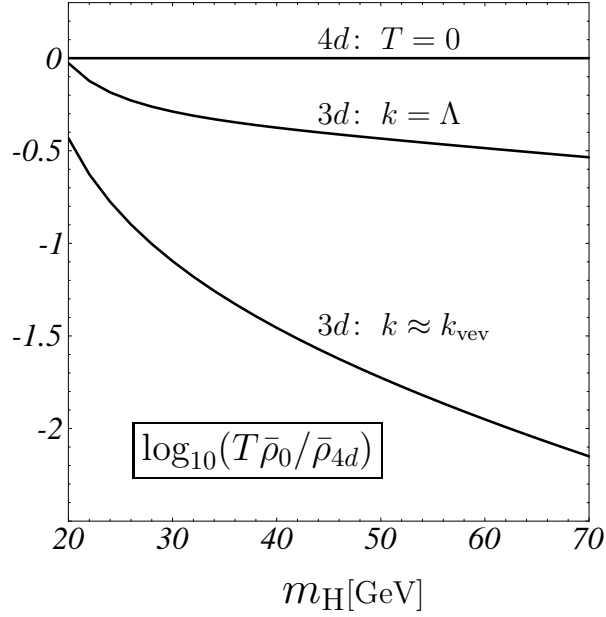,width=500\unitlength}
\end{picture}
\vskip-1.cm 
\begin{minipage}{.8\hsize}
\caption{\label{vev2} \small The size of the v.e.v.~$\rb_0(k)$ 
at $T=0$, $k=\xi T$ and at $k\approx k_{\rm vev}$ (in units 
of the $4d$ v.e.v.~$\rb_{4d}$ at $T=0$).}
\end{minipage} 
\end{center}
\end{figure}

\begin{figure}[t]
\begin{center}
\unitlength0.001\hsize
\begin{picture}(700,600)
\put(300,470){
\begin{tabular}{ll}
${\eb2=\ {\rm const.}}$&$  {}^{\put(0,0){\line(70,0){70}}}${}
\\[-.7ex]
${e_\star=\sqrt{6}\pi}$&${}^{\multiput(0,0)(3,0){24}{\line(0,0){1}}}${}
\\[-.7ex] 
${e_\star=\ 4}$&${}^{\multiput(0,0)(10,0){7}{\line(5,0){5}}}${}
\\[-.7ex] 
${e_\star=\ 3}$&${}^{\multiput(0,0)(20,0){4}{\line(10,0){10}}} $
\\[-.7ex] 
${e_\star=\ 2}$&${}^{\multiput(0,0)(28,0){3}{\put(0,0){\line(14,0){14}}}}${} 
\end{tabular}}
\put(250,85){ {\Large $m_{\rm H}$}[GeV]}
\put(120,200){ \fbox{{\huge $\log_{10}(\rb_0/T)$}}}
\hskip.04\hsize
\psfig{file=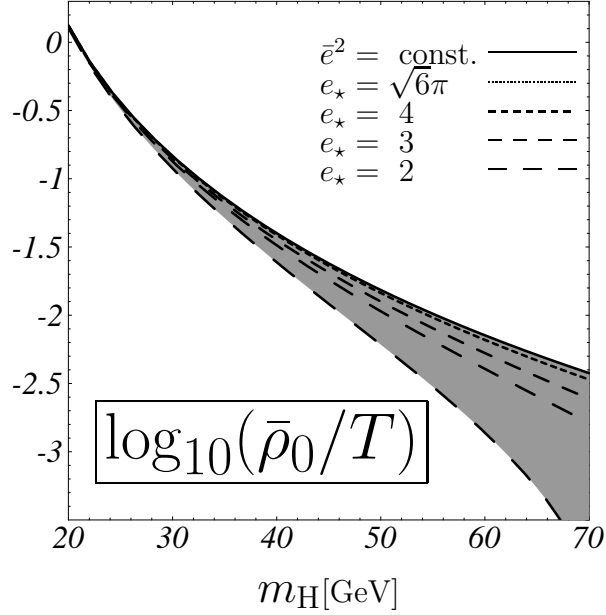,width=500\unitlength}
\end{picture}
\vskip-1.cm 
\begin{minipage}{.8\hsize}
\caption{\label{vevEs} \small The $3d$ v.e.v.~$\rb_0$ as
a function of the Abelian fixed point. }
\end{minipage} 
\end{center}
\end{figure}

\begin{figure}[t]
\begin{center}
\unitlength0.001\hsize
\begin{picture}(700,600)
\put(370,320){\fbox{\huge  $\sqrt{2}\phi_0/v$}}
\put(320,85){ {\Large $m_{\rm H}$}[GeV]}
\put(350,450){
\begin{tabular}{ll}
${e^3,\la^{3/2}}$
&${}^{\multiput(0,0)(20,0){3}{\put(0,0){\line(10,0){10}}
\put(14,0){\line(2,0){2}}}\put(60,0){\line(10,0){10}}}${}
\\[-1.ex]
${e^4, \la^2}$
&$  {}^{\put(0,0){\line(70,0){70}}}${}
\\
${e_\star=\sqrt{6} \pi}$
&${}^{\multiput(0,0)(3,0){24}{\line(0,0){1}}}${}
\\[-1.ex] 
${e_\star=\ 4}$
&${}^{\multiput(0,0)(10,0){7}{\line(5,0){5}}}${}
\\[-1.ex]
${e_\star=\ 2}$
&$ {}^{\multiput(0,0)(20,0){4}{\line(10,0){10}}} $
\end{tabular}}
\hskip100\unitlength
\psfig{file=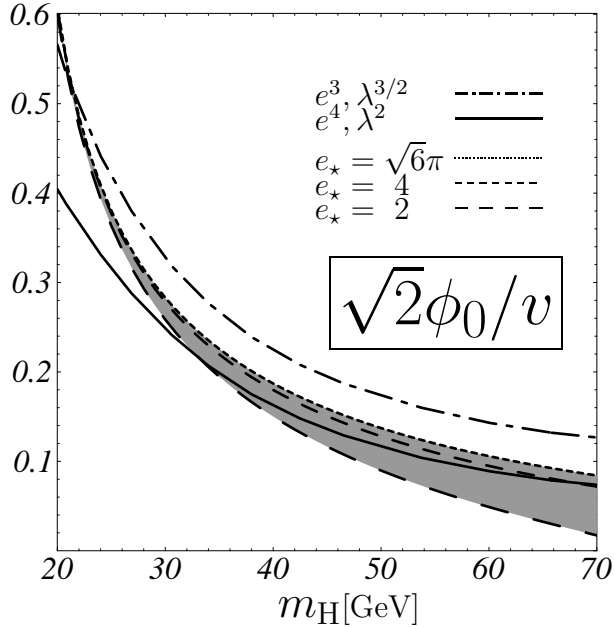,width=500\unitlength}
\end{picture}
\vskip-1cm
\begin{minipage}{.8\hsize}
\caption{\label{vev} \small  The v.e.v.~for various values of the 
Abelian fixed point in comparison with perturbation theory to order
$(e^3,\la^{3/2})$ and $(e^4, \la^2)$.}
\end{minipage} 
\end{center}
\end{figure}

\noindent
We begin with the discontinuity and the critical temperature, which
follow directly from solving the criticality conditions \eq{crit}.
\begin{figure}
\begin{center}
\unitlength0.001\hsize
\begin{picture}(700,600)
\put(170,430){\fbox{\Large ${T_c}/{m_{\rm H}}$}}
\put(350,95){{\Large $e_\star$}}
\put(350,510){{$m_{\rm H}=30$\,\,GeV}}
\put(350,360){{$m_{\rm H}=50$\,\,GeV}}
\put(350,210){{$m_{\rm H}=70$\,\,GeV}}
\hskip100\unitlength
\psfig{file=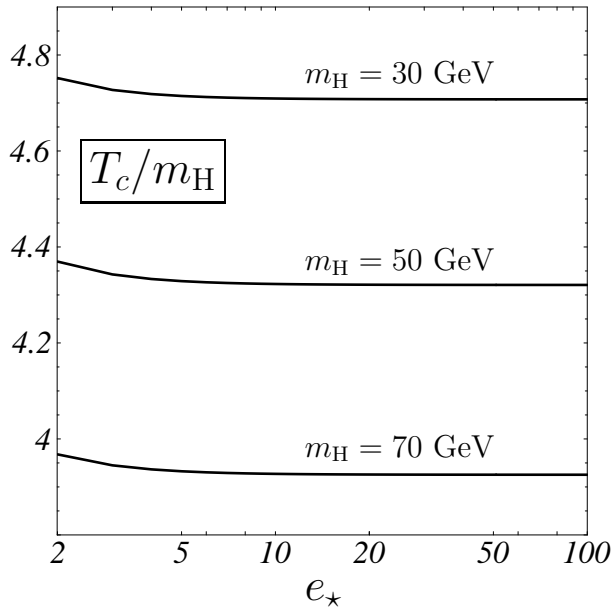,width=500\unitlength}
\end{picture}
\vskip-1.5cm
\begin{minipage}{.8\hsize}
\caption{\label{ptc}\small The critical temperature as a function of 
the Abelian fixed point.}
\end{minipage} 
\end{center}
\end{figure}
\noindent
The critical temperature as a function of the Abelian fixed point is 
given in Fig.~\ref{ptc} for $m_{\rm H}=30,50,70$\,GeV. 
It turns out that $T_c$ 
is rather insensitive against $\es2$. We observe an effect of a few 
percent only for very small values of $\es2$ 
(see also Fig.~\ref{varianceEs}). 
This is not a feature of the Higgs mass being relatively small, as 
similar results are obtained for all $m_{\rm H}$.
\\[1ex]
Before continuing, let us briefly compare our findings for the 
critical temperature to existing lattice data. Lattice results 
have been reported for $e=\s013$, $m_{\rm W}=80.6$ GeV and 
$m_{\rm H}=30$ GeV for the non-compact $U(1)$-Higgs model 
in \cite{Dimopoulos:1998cz}, and for the compact one 
in \cite{lattice}. The result reported in \cite{Dimopoulos:1998cz} 
is $T_c=131.18$ GeV for a finite lattice spacing. The continuum 
limit gives the slightly lower value $T_c=130.86$ GeV \cite{lattice}. 
This is consistent with $T_c=131.28$ GeV, the result for the compact 
case  \cite{lattice}. Here, for $\es2=6\pi^2$, we 
find $T_c=128.11$ GeV. As follows from Fig.~\ref{ptc}, 
the critical temperature is essentially independent of the 
effective Abelian fixed point. The perturbative value 
is $T_c=132.64$ GeV \cite{Dimopoulos:1998cz}. These results 
are in good numerical agreement.  
\\[1ex]
We now turn to the discussion of the discontinuity. In Fig.~\ref{vev2} we 
compare the logarithm 
of the v.e.v.~in $4d$
units (normalised to the v.e.v.~at $T=0$) at different scales. 
The renormalisation of
$\rb_0$ between the $T=0$ and the $k=\La$ lines results from  the
integration of the heavy and super-heavy modes, given by \eq{rhoR}.
The scale $k_{\rm vev}$ is defined as the scale where the running of
the potential minimum stops. This scale is related to the scale $k_M$, 
where the photon mass in the SSB regime is becoming larger than the coarse
graining scale, and thus decouples. Indeed, in the present approximation,
the flow equation for the potential minimum reads
\beq\label{runningRho}
\0{d\rb_0}{dk}=\01{\pi^2}\,\0{\eb2(k)}{\bar\la(k)}\,\ell^3_1(M^2(k)/k^2)\ .
\eeq 
Here, $\bar\la(k)=U''_k(\rb_0(k))$ denotes the quartic coupling at the 
minimum, and 
\beq\label{photonM}
M^2(k)=2\eb2(k)\rb_0(k) 
\eeq the photon mass squared. The running
of the v.e.v.~decouples at $k\approx k_{\rm vev}$, which happens as soon as 
the $3d$ photon mass $M$ is sufficiently larger than the scale $k$
(roughly at 
$M^2/k^2\approx 10$) such that the threshold function in \eq{runningRho} 
suppresses any further renormalisation ($k_{\rm vev}/T$ is displayed
in Fig.~\ref{scales}).
\\[1ex]
{From} Fig.~\ref{vev2} we conclude that the main part of the actual running 
of the potential minimum comes 
from integrating-out the $3d$ fluctuations, as can be
inferred from the wide separation of the $k=\La$ and the 
$k\approx k_{\rm vev}$ lines as opposed to the comparatively narrow separation
of the $T=0$ and the $k=\La$ lines. 
\\[1ex]
Fig.~\ref{vevEs} shows the  v.e.v.~$\rb_0$ as a function of the Higgs field
mass and the Abelian fixed point. The shaded region covers the region 
$2\le e_\star\le\infty$ for the Abelian fixed point value. For small 
$m_{\rm H}$, the effect is
clearly negligible. With increasing $m_{\rm H}$, however, the influence
of the running gauge coupling is increasing drastically, leading to a
strong weakening of the phase transition (see also Fig.\ref{varianceEs}).
\\[1ex]
\noindent
\vskip-1cm
\noindent
Finally we compare in Fig.~\ref{vev} the ratio of the $4d$ v.e.v.~$\phi_0$ 
to the initial $T=0$ v.e.v.~$v/\sqrt{2}$ for different Abelian fixed 
point values with the findings of perturbation theory.\footnote{We thank 
A.~Hebecker for providing his data from \cite{Hebecker93}  for comparison
in Figs.~\ref{vev}, \ref{Ucrit}, \ref{psigma} and \ref{latentheat}.}  
Again, the shaded region covers the region 
$2\le e_\star\le\infty$ for the Abelian fixed point. We observe
that the v.e.v.~shows a small dependence on the Abelian fixed point for
sufficiently small Higgs field mass. For larger values of 
$m_{\rm H}$, the v.e.v.~approaches
the perturbative two-loop result. It follows
that the v.e.v.~is  rather stable against effects from
the running Abelian charge, say a least for $\es2>20$. Only for 
$\es2\approx 4$ the running
becomes strong enough to result in a significant decrease of
$\rb_0$. 
\noindent

\subsection{The critical potential}
\begin{figure}
\begin{center}
\unitlength0.001\hsize
\begin{picture}(700,600)
\put(330,70){ \Large $\rb/T$}
\put(170,500){\fbox{\huge  $10^6 U_{\rm crit}/T^3$}}
\put(170,410){
\begin{tabular}{ll}
${e_\star=\ 100}$&$  {}^{\put(0,0){\line(70,0){70}}}${}\\[-.5ex]
${e_\star=\sqrt{6} \pi}$&$ 
{}^{\multiput(0,0)(10,0){7}{\line(5,0){5}}}${}\\[-1.ex]
${e_\star=\ 4}$&$ {}^{\multiput(0,0)(20,0){4}{\line(10,0){10}}} $\\[-1.ex]
${e_\star=\ 2}$&$
{}^{\multiput(0,0)(20,0){3}{\put(0,0){\line(10,0){10}}\put(14,0)
{\line(2,0){2}}}\put(60,0){\line(10,0){10}}}${}
\end{tabular}}
\hskip100\unitlength
\psfig{file=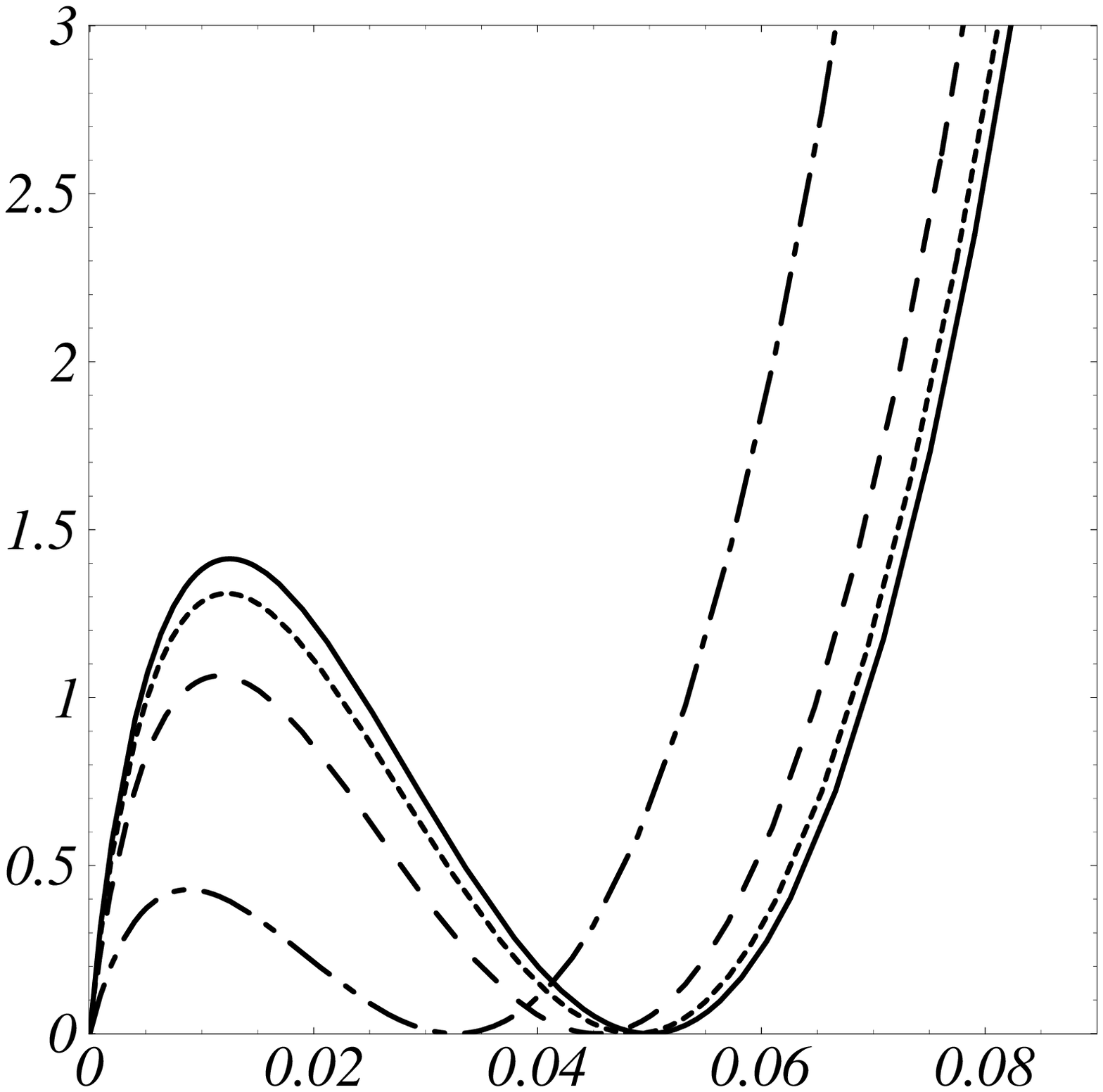,width=500\unitlength}
\end{picture}
\vskip-1.cm
\begin{minipage}{.8\hsize}
\caption{\label{Ucrites2} \small  The critical potential for different
values of the effective fixed point.}
\end{minipage} 
\end{center}
\end{figure}

\begin{figure}
\begin{center}
\unitlength0.001\hsize
\begin{picture}(700,600)
\put(300,50){ \Large $\sqrt{2}\phi/v$}
\put(150,510){\fbox{\huge  $10^7 T\ U_{\rm crit}/v^4$}}
\put(250,410){$a$}
\put(250,320){$b$}
\put(250,205){$c$}
\put(250,150){$d$}
\hskip100\unitlength
\psfig{file=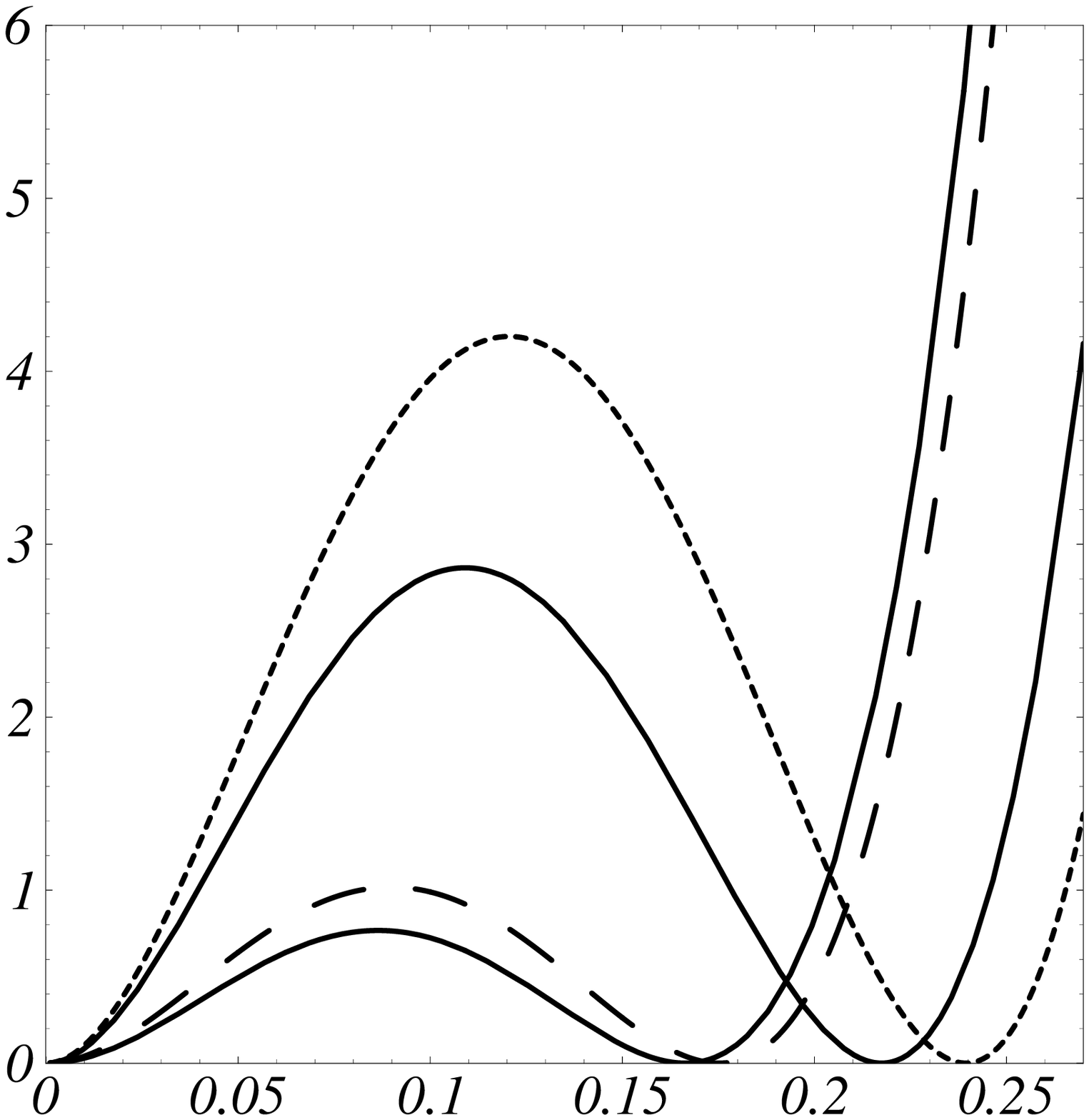,width=500\unitlength}
\end{picture}
\vskip-.5cm
\begin{minipage}{.9\hsize}
\caption{\label{Ucrit} \small  The critical potential, comparison with
perturbation theory (see text).}
\end{minipage} 
\end{center}
\end{figure}

\noindent
The critical potential is shown in Figs.~\ref{Ucrites2} and \ref{Ucrit} 
for $m_{\rm H} = 38$ GeV. Fig.~\ref{Ucrites2} gives the critical 
potential in units of the critical temperature
for different values of $\es2$ as functions of $\rb/T$.\footnote{Notice 
that comparing critical potentials (or other relevant quantities)
in units of $T$ for different values of $\es2$ is sensible due to the very
weak dependence of $T_c$ on the effective fixed point (see Figs.~\ref{ptc} 
and \ref{varianceEs}).}
We note that for large $\es2>6\pi^2$, the shape of the potential 
is rather insensitive against a change in $\es2$. Here, the additional 
scale dependence induced through the gauge coupling is quite small 
(a few percent). For small values of $\es2$, the height of the barrier 
is reduced significantly, up to a factor of 3 at $\es2=4$. The strong 
scaling of $\eb2$ thus weakens  the phase transition
considerably for small $\es2\ll 6 \pi^2$.  Again, the quantitative
change depends strongly on the value for the effective Abelian fixed
point, if $\es2 \ll 6\pi^2$. The non-trivial running of $\eb2(k)$ has
a stronger effect on the small $\rb$ region of the potential. Here, the 
decoupling of the gauge field sets in only at smaller scales, which in
turn results in a stronger quantitative effect due to the running
gauge coupling. 
\noindent
\noindent
Fig.~\ref{Ucrit} gives the critical potential in units of 
the $4d$ v.e.v.~$v/\sqrt{2}$, and
compares the solution of \eq{flowpot1} with those obtained within
perturbation theory (PT). Line (a) corresponds to PT to 
order $(e^3,\la^{3/2})$ \cite{PT1}, line (b) to our result 
with $\es2=6\pi^2$, 
line (c) to PT at order $(e^4,\la^2)$ \cite{Hebecker93} and
line (d) to our result with 
$\es2=4$. For $\es2=6\pi^2$,
the critical potential is situated half way between the one- and 
two-loop perturbative results. For decreasing $\es2$, the critical 
potential approaches quickly the two-loop result, and becomes even 
smaller at about $\es2\approx 4$. It is interesting to note that 
a value for $\es2$ can be found for which the two-loop perturbative 
result is matched perfectly.

\subsection{Surface tension and latent heat}
\begin{figure}[t]
\begin{center}
\unitlength0.001\hsize
\begin{picture}(700,600)
\put(350,500){\begin{tabular}{ll}
${e^3,\la^{3/2}}$
&${}^{\multiput(0,0)(20,0){3}{\put(0,0){\line(10,0){10}}
\put(14,0){\line(2,0){2}}}\put(60,0){\line(10,0){10}}}${}
\\[-1.ex]
${e^4, \la^2}$
&$  {}^{\put(0,0){\line(70,0){70}}}${}
\\
${e_\star=\sqrt{6} \pi}$
&${}^{\multiput(0,0)(3,0){24}{\line(0,0){1}}}${}
\\[-1.ex] 
${e_\star=\ 4}$
&${}^{\multiput(0,0)(10,0){7}{\line(5,0){5}}}${}
\\[-1.ex]
${e_\star=\ 2}$
&$ {}^{\multiput(0,0)(20,0){4}{\line(10,0){10}}} $
\end{tabular}}
\put(200,200){\fbox{{\huge $\log_{10}(\sigma)$}}}
\put(300,70){ {\Large $m_{\rm H}$}[GeV]}
\hskip100\unitlength
\psfig{file=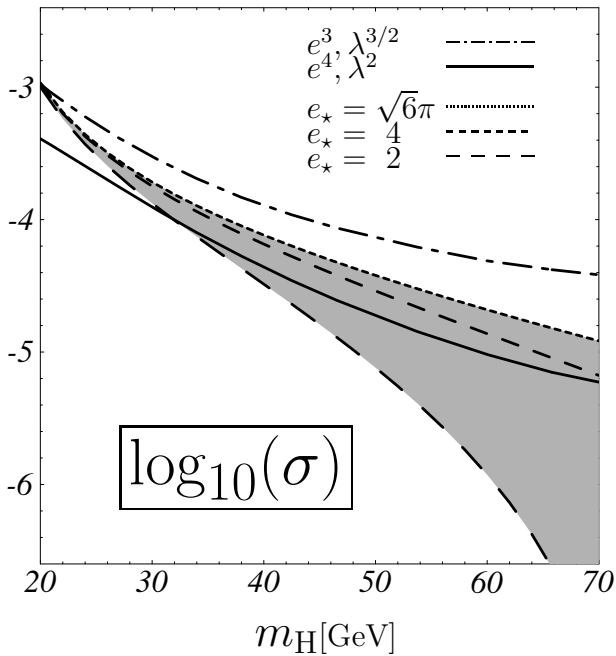,width=500\unitlength}
\end{picture}
\vskip-.5cm
\begin{minipage}{.8\hsize}
\caption{\label{psigma} \small The surface tension for various values of the 
Abelian fixed point in comparison with perturbation theory to order
$(e^3,\la^{3/2})$ and $(e^4, \la^2)$.}
\end{minipage} 
\end{center}
\end{figure}

\noindent
The interface tension for a planar interface separating the two
degenerate vacua follows from \eq{ansatz} as
\beq\label{sigma}
\sigma=2 \int_0^{\F_+} d\F \sqrt{ Z_\F U_{\rm crit}(\rb)}\ .
\eeq
It is sensible to the actual shape of the critical potential and yields
additional information regarding the strength of the phase transition. 
In Fig.~\ref{psigma} the surface tension is shown as a function 
of $m_{\rm H}$ and in comparison with perturbation theory. The shaded region
covers the results for $2\le e_\star\le \sqrt{6}\pi$. We again note that the
effect of the running coupling is negligible for small Higgs boson
mass. In contrast to the v.e.v., the surface tension depends rather
strongly on $\es2$ already for moderate values of $m_{\rm H}$. 
An even stronger running of $e^2_3$ would lead to a dramatic
decrease of the surface tension, up to several orders of magnitude. 
\\[1ex]
\noindent
Finally, we consider the latent heat $L$, defined at the critical 
temperature as 
\beq\label{latentheat1}
L = T \left.\left(\0{dU(\rb_0)}{dT}~-~\0{dU(0)}{dT}\right)\right|_{T=T_c}
\eeq
Using \Es{crit}, \eq{F} and \eq{initialpot}  we obtain
\beq\label{latentheat2}
{L}=  \left({m_{\rm H}^2} - 2{m_{\rm R}^2}\right) \rb_0 
     + \012 \la_{\rm R} T\rb_0^2 
     +   3{\De}(\rb_0)
     - \rb_0{\De}'(\rb_0)
\eeq
The latent heat is related to the discontinuity and the 
mass of the scalar particle. Using \eq{crit}, it can be shown that
\beq\label{clausius}
L=\rb_0\ m_{\rm H}^2\,,
\eeq
which is also known as the Clausius-Clapeyron equation. This 
relation was shown to be fulfilled within an
explicit gauge-invariant perturbative calculation \cite{GI}. However, it
holds not true within standard perturbation theory: the perturbative 
values for the latent heat 
as found in \cite{Hebecker93} are all below the value given through the 
Clausius-Clapeyron relation \eq{clausius}. The deviation varies between 
a few percent up to 15-20$\%$ for $m_{\rm H}$ between 20~GeV and 70~GeV, 
and is larger at order $(e^4,\lambda^{2})$ than at 
order $(e^3,\lambda^{3/2})$.
\begin{figure}[t]
\begin{center}
\unitlength0.001\hsize
\begin{picture}(700,600)
\put(350,500){\begin{tabular}{ll}
${e^3,\la^{3/2}}$
&${}^{\multiput(0,0)(20,0){3}{\put(0,0){\line(10,0){10}}
\put(14,0){\line(2,0){2}}}\put(60,0){\line(10,0){10}}}${}
\\[-1.ex]
${e^4, \la^2}$
&$  {}^{\put(0,0){\line(70,0){70}}}${}
\\
${e_\star=\sqrt{6} \pi}$
&${}^{\multiput(0,0)(3,0){24}{\line(0,0){1}}}${}
\\[-1.ex] 
${e_\star=\ 4}$
&${}^{\multiput(0,0)(10,0){7}{\line(5,0){5}}}${}
\\[-1.ex]
${e_\star=\ 2}$
&$ {}^{\multiput(0,0)(20,0){4}{\line(10,0){10}}} $
\end{tabular}}
\put(150,200){\fbox{{\huge $\log_{10}(L/T^3)$}}}
\put(300,70){ {\Large $m_{\rm H}$}[GeV]}
\hskip100\unitlength
\psfig{file=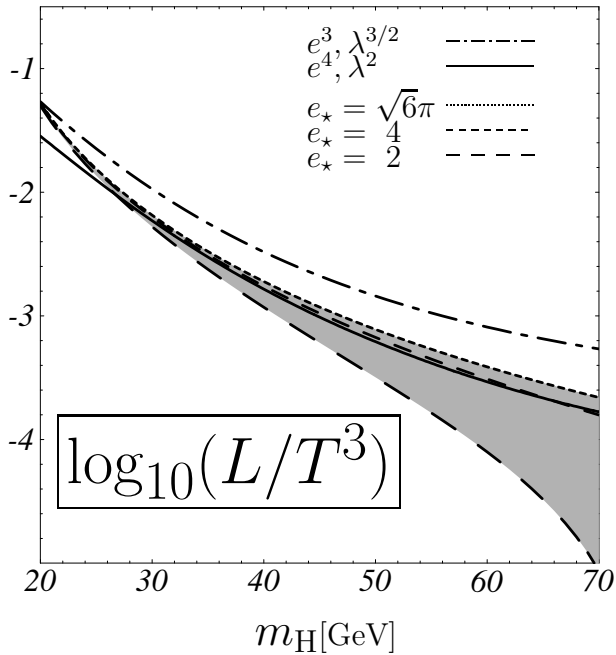,width=500\unitlength}
\end{picture}
\vskip-.5cm
\begin{minipage}{.8\hsize}
\caption{\label{latentheat} \small The latent heat for various values of the 
Abelian fixed point in comparison with perturbation theory to order
$(e^3,\la^{3/2})$ and $(e^4, \la^2)$.}
\end{minipage} 
\end{center}
\end{figure}
\noindent
The latent heat in units of the critical temperature is displayed in
Fig.~\ref{latentheat} for various values of the effective Abelian fixed 
point, and in comparison with perturbation theory to order 
$(e^3,\lambda^{3/2})$ and $(e^4,\lambda^{2})$. 
The shaded region covers the interval 
$2\le e_\star\le \sqrt{6}\pi$. We again observe a sharp decrease for small
$e_\star$ and large Higgs boson mass as in Fig.~\ref{vev}. It is 
interesting to note that the curve for $e_\star=4$ roughly agrees with 
the two-loop perturbative result for all $m_{\rm H}$ above 30~GeV. This is 
not the case for the surface tension. Comparing 
Fig.~\ref{latentheat} with Fig.~\ref{psigma}, we notice that the effect of
the running gauge coupling is more pronounced for the surface tension, because
the entire region for $\rb\le\rb_0$ enters \eq{sigma}, while the latent heat 
is only affected by $\rb_0$.

\subsection{Characteristic scales}\label{char-scales}
\begin{figure}
\begin{center}
\unitlength0.001\hsize
\begin{picture}(700,700)
\put(300,10){{\Large $m_{\rm H}$} [{GeV}]}
\put(170,525){$\La=\xi T$}
\put(440,315){$\k-{cr}$\ {\footnotesize $(\es2=4)$}}
\put(200,400){$k_s$}
\put(220,330){$k_{\rm vev}$}
\put(265,265){$k_{\rm stable}$}
\put(170,245){$k_{\rm flat}$}
\put(150,160){$\k-{cr}\ ${\footnotesize $(\es2={6}\pi^2)$}}
\put(360,580){\fbox{{\Large  $\log_{10}(k/T)$}}}
\hskip100\unitlength
\psfig{file=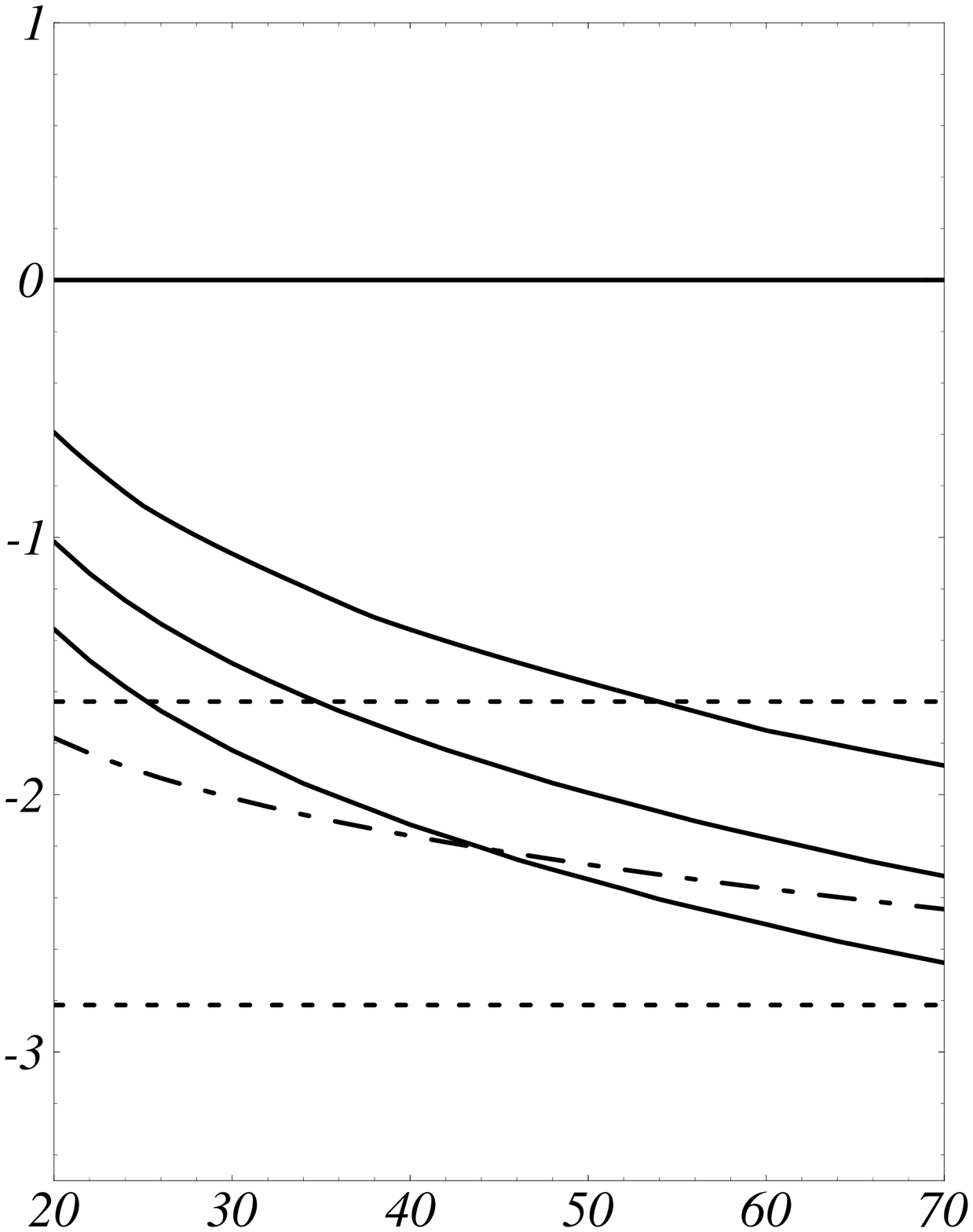,width=500\unitlength}
\end{picture}
\begin{minipage}{.8\hsize}
\caption{\label{scales} \small  Characteristic scales (see text).}
\end{minipage} 
\end{center}
\end{figure}

\noindent
We discuss the results obtained so far in terms 
of the characteristic scales relevant for the phase transition. Most 
of the qualitative (and even quantitative) features can be understood 
once these scales are known.
\\[1ex]
In Fig.~\ref{scales}, we have depicted the relevant momentum scales as
a function of the Higgs mass. The top line at $k=\La$ corresponds to
the scale of dimensional reduction, that is, the starting point of
the flow in $3d$. The scales $k_s,k_{\rm vev}$ and $k_{\rm stable}$
(full lines) describe characteristics of the potential, the scale
$\k-{cr}$ (dashed lines, for two values of the Abelian fixed point)
the characteristics of the gauge sector, and $k_{\rm flat}$ 
(dashed-dotted line) the scale where scalar fluctuations can no longer 
be neglected within the non-convex part of the potential. All these scales
are now discussed in detail. 
\\[1ex]
At $k=k_s$,
the origin of the effective potential stabilizes, $U'(\rb=0)=0$, as
the mass term squared at vanishing field changes sign. The free energy
has two local minima for scales below $k_s$. This scale is therefore a good
estimate for the scale of discontinuity.  In \cite{BFLLW}, an
estimate for this scale has been given, based on a
local polynomial approximation for the potential. Within our
conventions, it reads $k_{\rm dis}\approx 0.18 e^4(T)/\la_{\rm R}(T)$
for a sharp cut-off, and roughly coincides with $k_s$ as presented
here ($k_{\rm dis}/k_s$ ranges between 1 to 3).
\\[1ex]
The scale $k\approx k_{\rm vev}$ indicates when the v.e.v.~$\rb_0$ is
within $1\%$ of its final value, eventually reached for $k\to
0$. However, this is not yet the scale where the critical potential
has reached a stable shape, which actually happens only at about
$k\approx k_{\rm stable}$. This results from the fact that the 
effective photon mass squared $2\eb2(k)\rb$ (within the non-convex 
part of the potential) is smaller than the photon mass at the minimum 
in the SSB regime \eq{photonM}, and the decoupling takes 
place only at smaller scales. Here, we have obtained $ k_{\rm stable}$
comparing the depth of the potential $U(0)-U(\rb_0)$ at $\rb_0$ with
the height of the barrier $U(\rb_{\rm max})-U(0)$, demanding this
ratio to be below $\approx$ 5\%. At $k =k_{\rm stable}$, the v.e.v.~is
as close as 0.1\% to its final value. \footnote{Remember that the critical 
potential at $k_{\rm stable}$, within the present approximations, is about
the same as at $k=0$, as no substantial running takes place below
$k_{\rm stable}$.}
\\[1ex]
The cross-over scale $\k-{cr}$ characterises the cross-over from the
Gaussian to the Abelian fixed point. For $\es2=6\pi^2$, we see that
$\k-{cr}$ is about 1-2 orders of magnitude smaller than the scale
$k_s$, which explains why the running gauge coupling has, in this
case, only a small numerical effect on the properties of the phase
transition. From the fact that the scales $k_{\rm vev}$ and $k_{\rm
stable}$ are separated by an order of magnitude ($k_{\rm vev}/k_{\rm
stable}\approx 5$), we can conclude that the running of the gauge
coupling has a stronger effect on physical observables based on the
entire effective potential (like the surface tension), than those
related only to the v.e.v.~(like the latent heat). This is
quantitatively confirmed by the findings displayed in the
Figs.~\ref{vevEs}, \ref{vev}, \ref{psigma} and \ref{latentheat}.  For
$\es2=4$, we realize that the corresponding cross-over scale is of the
same order of magnitude as the scales $k_s,\ k_{\rm vev}$ and $k_{\rm
stable}$.\footnote{In Fig.~\ref{scales}, the scales $k_s,\ k_{\rm vev}$, 
$k_{\rm stable}$ and $k_{\rm flat}$ have been obtained for $\es2=6\pi^2$. 
The corresponding results for $\es2=4$ do deviate (for larger Higgs mass) 
only slightly from the curves as presented here. This minor difference is
of no relevance for the present discussion.}  
This is the region where the running of the gauge coupling
has a strong quantitative effect on the properties of the phase
transition, leading to a significant decrease of the strength of the
transition.
\\[1ex]
\noindent
Finally, we have also indicated the scale $k_{\rm flat}$
(dashed-dotted line), which is an estimate for the scale where the
flattening of the inner part of the effective potential sets in. We
obtained $k_{\rm flat}$ from solving $k^2+U_k'(\rb)\approx 0$
numerically for $k$ in the non-convex part of the potential, with
$U_k$ the leading order solution for the free energy.\footnote{A
similar though slightly shifted curve for $k_{\rm flat}$ is obtained
from solving $k^2+U_k'(\rb)+2\rb U''_k(\rb)\approx 0$.}
\\[1ex]
In \cite{Litim97} an estimate for the ratio of $k_{\rm flat}/k_{\rm
stable}$ has been obtained, based on an investigation of the surface
tension of the $3d$ Abelian Higgs model in the universal limit
$\La\to\infty$. There, it was found that $k^2_{\rm flat}/k^2_{\rm
stable}\approx \eb2/M$, with $M$ being the $3d$ photon mass. The
boundary $k^2_{\rm flat}/k^2_{\rm stable}\approx 1$ yields the
relation $k_{\rm flat}\approx (e^2T/2\rb_0)^{1/4}k_{\rm stable}$,
which, using the data for $k_{\rm stable}$ as in Fig.~\ref{scales},
coincides within a few percent with the line for $k_{\rm flat}$ as
obtained above. Corrections to the universal limit can be expanded as
a series in $M^2/\La^2$ \cite{Litim97}. In the present case, we start
at a finite scale $\La=\xi T$, but the smallness of $M^2/\La^2$
(ranging from 0.2 to  0.001 for $20\,{\rm GeV}\le m_{\rm H}\le 70$\,GeV) 
is responsible for the small corrections with respect to the universal
limit $\La\to\infty$. Being close to the universal limit of the effective 
$3d$ theory also explains
why the dependence on the matching parameter $\xi$ is rather small.
\\[1ex]
We now come back to the discussion of $k_{\rm flat}$ from
Fig.~\ref{scales}, which, by definition, sets the scale below which
the scalar fluctuations trigger the flattening within the non-convex
part of the potential, and hence the scale below which these
fluctuations should no longer be neglected. First notice, that the
scale of discontinuity $k_s$ is bigger than  $k_{\rm flat}$ by an
order of magnitude. We can thus expect that the scale of discontinuity
is only weakly affected by the scalar fluctuations. Also, $k_{\rm
vev}>k_{\rm flat}$ by a factor of $\approx 5$. Finally,  for small
Higgs field mass, $k_{\rm flat}$ is also smaller than $k_{\rm
stable}$. In this region, only small quantitative changes are expected
if the scalar fluctuations are taken into account. This is no longer
the case for large Higgs field mass, where $k_{\rm flat}\ge k_{\rm
stable}$. However, as these effects concerns mainly the non-convex 
part of the potential, and thus quantities like the surface tension, 
we can still expect that the latent heat and the v.e.v.~are only 
moderately affected. 
\\[1ex]
These last observations are also relevant for the applicability of
Langer's theory of bubble nucleation. The concept of an interface
tension, as defined in \eq{sigma}, is based on the implicit assumption
that the scale $k_{\rm stable}$ can indeed be identified. A criterion
for this being the case is the smallness of the perturbative expansion
parameter. From our consideration we can conclude that this will
become more and more difficult for increasing $e^2T_c/2\rb_0\ge 1$,
that is, for very weakly first-order phase transitions.\footnote{The
treatment of very weakly first-order transitions based on
coarse grained potentials has been considered in \cite{tunneling}.}

\subsection{Higher order corrections}\label{error}
\noindent
Finally, we comment on the higher order
corrections which are expected from operators neglected within the present
approximation.  Clearly, the results presented here are affected
by the approximations performed,  most notably through $(i)$ the derivative
expansion, $(ii)$ neglecting the scalar field  fluctuations as opposed
to the gauge field ones, $(iii)$ approximating the infra-red regime
of the Abelian charge by an effective fixed point, and $(iv)$ computing
the initial conditions perturbatively. We discuss these
approximations now one by one.  
\\[1ex]  
$(i)$ The leading order terms of the
derivative expansion are known to correctly describe critical
equations of state and scaling solutions for a variety of $O(N)$-symmetric
scalar
models in $3d$. Although little is known about the convergence of such
an expansion, it appeared that the smallness of the anomalous dimensions
controls the influence of higher order derivative operators in the
effective action. Therefore, an {\it a posteriori} consistency check
for the reliability of the derivative  expansion consists in computing
the corresponding scalar and gauge field anomalous dimension
$\eta_\varphi$ and $\eta_F$. In the present case, this involves more
complicated higher order threshold functions (for their definitions
and further details, see \cite{BFLLW}). At the scale $k\approx k_{\rm
stable}$, we can compute the scalar anomalous dimension
self-consistently from the explicit solution for the effective
potential, obtained while neglecting $\eta_\varphi$. We find that
$|\eta_\varphi|\le 0.005$ in the interval considered, which is
consistent with our initial approximation $\eta_\varphi=0$ and
justifies the derivative expansion within the scalar sector. For
$N=1$, the gauge field anomalous dimension  $\eta_F$ can be estimated
in a similar way. It becomes of order one only when the non-trivial
fixed point is approached. We find that $\eta_F$ ranges from $0.03$ to
$0.4$ within the range of Higgs field masses considered here and for
$\es2\approx 6\pi^2$. A
main difference between the scalar and the gauge field sector is that
the gauge field anomalous dimension grows large ($\eta_F=1$) at a
scaling solution. Therefore, one expects that higher order corrections
within a derivative expansion (or the momentum dependence of the gauge
coupling) can become important at a scaling solution and should not be
neglected. In the present case, however, the scales relevant for the
first-order phase transition have been reached before the Abelian
charge finally runs into its non-trivial fixed point, that is before
$\eta_F=1$. Therefore we can expect that the derivative expansion behaves 
reasonably well even for the gauge field sector.
\\[1ex]  
$(ii)$ In the same way, we can check the validity of neglecting
scalar fluctuations within the non-convex part of the effective
potential. It is found that the self-consistent inclusion of scalar
fluctuations to leading order results in corrections of the order of a
few percent, increasing with increasing Higgs field mass (see
Appendix \ref{AppC}). This agrees also with the discussion of the
preceding section, where it was argued that scalar fluctuations should
no longer be neglected as soon as $k_{\rm flat}$ is of the order of
$k_{\rm stable}$. Clearly, the weaker the first-order phase transition
the more scalar fluctuations will become relevant at the phase
transition.   For a quantitatively more reliable computation of
thermodynamical quantities in the weakly type-I 
region, one has to go beyond the
present approximation and include scalar fluctuations. All
the present approximations can be improved in a systematic way, as
has been emphasized earlier. This can be
done either along the lines outlined in Sect.~\ref{runningp}, or by a
straightforward numerical integration of the flow equation as in
\cite{Tetradis97}.  
\\[1ex]  
$(iii)$ The main uncertainty in the present
understanding of the $U(1)$-Higgs theory is linked to the gauge sector
of the theory {\it i.e.}~the precise infra-red behaviour of the Abelian
gauge coupling.  Here we have effectively parametrised this
uncertainty in terms of an Abelian fixed point motivated by previous
work based on large\,-$N$ extrapolations and Wilsonian RG techniques. A
precise determination of the correct fixed point requires the study of
the momentum and of the field dependence of the Abelian charge. Our
approximation assumes that the field gradients of the function
$\es2(k,\rb)$ remain sufficiently small within the non-convex part of
the potential at scales above $k\approx k_{\rm stop}$. In the
large\,-$N$ limit, where this fixed point is well understood, the
results in the present approximation are in very good agreement with
the result found within a fixed dimension computation.  
\\[1ex]   
$(iv)$ The points $(i)-(iii)$ concerned the approximations on the level 
of the flow equation. These are the most important ones, because they act 
back on $\Gamma_k$ upon integration of the flow. An additional
approximation concerns the initial conditions to the flow. 
Here, they have been obtained from the 
dimensional reduction scenario within a perturbative loop computation. For 
the present purposes, it was sufficient to use a 1-loop perturbative matching 
as given in Sect.~\ref{initial}, although the 2-loop matching has 
been reported as well \cite{K94}. These higher order effects can be taken 
into account in principle; in practice, this shall not be necessary 
because their quantitative influence is smaller than the contributions 
from the scalar fluctuations for larger Higgs field mass,
which have already been neglected. In 
any case, a small change of the initial condition cannot change 
the main effect reported here. Except for small Higgs field masses, 
the dominant contributions 
come from integrating-out modes in $3d$. This follows directly
from  Fig.~\ref{vev2}, 
which shows that the main running of the v.e.v.~takes place below 
the scale of dimensional reduction. 
\\[1ex] 
Finally, we remark that the quality of a given
approximation can also be assessed by studying the dependence on the
coarse graining scheme. This discussion will be the subject of the
following section.
\section{Scheme dependence}\label{schemes}
\noindent
All quantitative results present up to now have been obtained for a
sharp cut-off regulator. It is a straightforward consequence of the 
Wilsonian renormalisation group approach 
that physical observables obtained from a solution to a 
Wilsonian flow equation will not depend on the precise form of the 
coarse-graining. Unfortunately, this conclusion holds only if the 
$\it full$ effective action is computed. On a technical level, this 
is barely possible, and truncations of the effective action have to be 
employed. It is precisely this truncation that can introduce a spurious 
coarse graining scheme dependence for physical observables. In this 
section we address the question as to
what extend the physical observables as obtained in the preceding
section do (or do not) depend on the precise form of the coarse
graining. In doing so, we are able to present quantitative `error bars'
related to the scheme dependence. We also present evidence for an intimate
quantitative link between the scheme dependence and the truncations employed.

\subsection{Scheme dependence vs. truncations}\label{quali}
\noindent
Consider the case of computing some physical observable from the solution
to a (truncated) Wilsonian flow. It goes without saying that a {\it strong} 
dependence of this observable on the coarse graining employed is not 
acceptable as it would cast serious doubts on the truncations 
performed so far. With `strong' we mean `inducing large quantitative', 
or even `qualitative' changes.
On the other hand, a {\it weak} scheme
dependence of certain physical observables is a sign for the viability
of the approximation employed. In fact, if we 
were able to solve the flow equations without truncating the
effective action $\Gamma_k$, the final result in the physical limit
$k\to 0$, which is by construction nothing else but the full quantum
effective action $\Gamma$, should not depend on the details of the
particular coarse graining employed. There is little hope for this
holding true for any truncation of the effective action
$\Gamma_k$ as any truncation necessarily neglects infinitely
many operators.
\\[1ex]
The coarse graining procedure is implemented through the
momentum-dependent operator $R_k(q^2)$. It couples to all the operators
present in $\Ga_k$ in a well-defined way, that is, 
according to the flow equation
\eq{general}. Replacing a coarse-graining by another
coarse graining implies that the effective coupling of $R_k(q^2)$
to the operators contained in the effective action changes
accordingly. A truncation of the effective action amounts to neglecting
infinitely many operators to which the coarse graining, in principle,
is sensitive. Therefore, studying the scheme dependence will probe 
whether some relevant operators (for the problem under
investigation) have been neglected, or not. In this light,
the indirect feed-back of some relevant operators should manifest
itself through some strong eigenmode with respect to a change of the coarse
graining procedure.
\\[1ex]
Although these arguments, as presented so far, are of a purely
qualitative nature, we will show in the
sequel  that they can indeed be given a quantitative
meaning.

\subsection{Coarse grainings}\label{regulator}
\noindent
Before studying in detail the scheme dependence of our results, we
will briefly review the main requirements for a viable coarse graining
procedure. There are basically three key points to be considered. The
first one concerns the possible zero-modes of the propagators, which
typically cause strong infra-red problems within perturbative loop
expansions in $d<4$ dimensions. These are properly regularised, if
\begin{mathletters}\label{ConditionsRk}
\beq
\label{1}  \lim_{q^2\to0}\,     R_k(q^2)>0
\eeq
holds true. This way, the effective inverse propagator for a massless 
mode reads $q^2+R_k(q^2)$, and has a well-defined infra-red limit. 
The second point concerns the infra-red limit of the effective 
action $\Gamma_k$, which should coincide with the usual effective 
action for $k\to 0$. This is the case, if
\beq
\label{0}  \lim_{k\to 0}\,      R_k(q^2)=0\ .
\eeq
Finally, we have to make sure that the correct initial effective 
action in the ultra-violet limit is approached which is guaranteed by
\beq
\label{00} \lim_{k\to\infty}\,  R_k(q^2)\to \infty\ .
\eeq
\end{mathletters}%
Any function $R_k(q^2)$ with the above properties can be considered 
as a coarse graining \cite{Wetterich91,Litim97}. It is convenient 
to re-write $R_k$ in terms of dimensionless functions $r(q^2/k^2)$ as
\beq
R_k(q)=Z\, q^2\, r(q^2/k^2)\ ,
\eeq
where $Z$ corresponds to a possible wave-function renormalisation 
($Z_\phi=1$ in our approximation).
\\[1ex]
Let us introduce two classes of regulator functions which are commonly 
used in the literature.  The first one 
is a class of {\it power-like} regularisation 
schemes given by the coarse-graining function
\beq\label{power}
r_p(y)=y^{-n} \ ,
\eeq
and $y\equiv q^2/k^2$. The particular case $n = 1$ corresponds to a 
mass-like regulator 
$R_k\sim k^2$, and $n = 2$ to a quartic regulator $R_k\sim k^4/q^2$. 
These algebraic regulators are often used because the related 
threshold functions can be computed analytically. On the other 
hand, these regulators decay only algebraically for large momenta, 
which can in principle lead to an insufficiency in the integrating-out 
of the hard UV modes.  
\\[1ex]
A second convenient class of regulators consists of exponential ones, 
parametrised as
\beq\label{exp}
r_e(y) = \01{\exp(c\,y^{n})-1}\, ,
\eeq
where $c$ is a constant. The exponential regulator with $n=c=1$ 
has been used previously in various numerical investigations 
\cite{BFLLW,BLLW,Litim94a}. The suppression of large momentum modes 
$q^2\gg k^2$ to the flow is now exponential and thus much stronger than 
in the case of algebraic regulators. It is expected that this property 
is at the basis for a good convergence of approximate solutions.
\\[1ex]
Both classes of regulator functions depend on the parameter $n$, with 
$1\le n\le \infty$. In the limit $n\to \infty$, they both approach to 
what is known as the sharp cut-off regulator, given 
by \cite{Wegner73,Polchinski84}
\beq
r_s(y)=\01{\theta(y-1)}-1\, \label{sharp}\ .
\eeq
We will now consider the dependence of certain physical observables 
on particular choices of these regulators.
\subsection{Tri-critical point and large\,-$N$ limit}\label{tricrit}
\begin{figure}[t]
\begin{center}
\unitlength0.001\hsize
\begin{picture}(500,530)
\put(270,165){{\Large $n$}}
\psfig{file=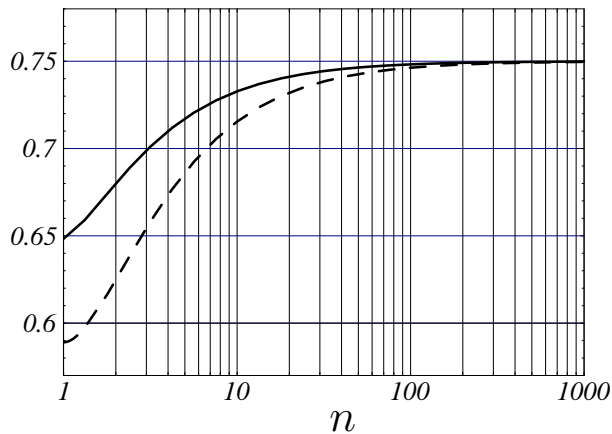,width=500\unitlength}
\end{picture}
\vskip-2.5cm
\begin{minipage}{.8\hsize}
\caption{\label{a1} \small  The expansion coefficient $a_1$ (see text).}
\end{minipage} 
\end{center}
\end{figure}

\noindent
We have given an estimate for the endpoint of the critical line in 
\eq{mh-crit}.  Its mere existence is closely linked to the presence 
of an Abelian fixed point, although it will be within the domain of 
validity only for small values of the latter. Both functions $F_1$ 
and $F_2$ depend explicitly on the RS, and so does the solution 
to eqs.~\eq{crit}.  In the general case, the endpoint of the critical 
line also depends on the RS. Instead of \eq{mh-crit}, which is 
the result for a sharp cutoff, we find for the general case
\beq\label{endpoint}
\0{m_{\rm H}^2}{M^2}=\0{8a_1}{3\pi^2}\,\es2 \ ,
\eeq
where terms ${\cal O}(e)$ have been dropped. The entire scheme 
dependence is now encoded in the coefficient $a_1$, given by
\beq
a_1=-\032\int^\infty_0 dy\,\0{\,r'(y)\,y^{-\frac{1}{2}}}{[\,1+r(y)\,]^{3}}
\eeq
in $d=3$ dimensions.  This coefficient 
belongs to a set of expansion coefficients 
$a_k$ characterising a coarse graining scheme 
(see Appendix~\ref{AppA} for their general definition and more details).
For each of the two classes of regulators the coefficient $a_1$ can be
calculated as a function of the parameter $n$. In Fig.~\ref{a1},
the dashed line corresponds to the power-like, and the full line to 
the exponential regulator class with $c = \ln2$. For this choice of
$c$ both set of regulators are normalised to $r(1)=1$.
For a power-like regulator, we 
find explicitly $a_1=\s034\Gamma[1+\s01{2n}]\Gamma[2-\s01{2n}]$, and
for the exponential one $a_1=\s03{8}\,n^{-1}\,c^{1/2n}
(2^{1/2n}-2)\Gamma[-\s01{2n}]$. It is
interesting to note that although these classes of regulators
do have strong qualitative differences, the coefficient $a_1$, 
which only involves a folding of $r(y)$ over all momenta, is 
rather stable ({\it i.e.} $\pm 10\%$ about the mean value).
\\[1ex]
\noindent
We shall compare the numerical value of the tri-critical point with 
results obtained in the large\,-$N$ limit via the $\eps$-expansion 
\cite{Arnold1} or a fixed dimension 
computation in $(d=3)$ \cite{Arnold2}. As argued in Sect.~\ref{runninge}, 
the Abelian fixed point reads $\es2=6\pi^2/N$ in the large\,-$N$ limit, 
and our above result therefore becomes
\beq\label{tric}
\0{\la_{3}}{e^2_3}=16 a_1\01N\approx {(9.4 - 12.0)}\01{N}
\eeq
The $\eps$-expansion, to leading order, yields 
\beq\label{tricE}
\0{\la_{3}}{e^2_3}=(54-136\eps)\01N
\eeq
This is to be compared to the result of \cite{Arnold2}, which reads
\beq\label{tricA}
\0{\la_{3}}{e^2_3}=\0{96}{\pi^2}\01N\approx {9.9}\01{N}
\eeq
While \eq{tricE} fails to give a reliable answer at $\eps=1$, we observe 
that our result \eq{tric} is in good numerical agreement with \eq{tricA}.

\subsection{Scheme dependence of the critical potential}
\begin{figure}
\begin{center}
\unitlength0.001\hsize
\begin{picture}(1000,500)
\put(670,-20){\large $\sqrt{2}\phi/v$}
\put(270,-20){\large $\sqrt{2}\phi/v$}
\put(650,400){\fbox{\Large  $10^8 T\ U_{\rm crit}/v^4$}}
\put(200,400){\fbox{\Large  $10^7 T\ U_{\rm crit}/v^4$}}
\put(170,90){{$m_{\rm H}=38$\,\,GeV}}
\put(620,90){{$m_{\rm H}=70$\,\,GeV}}
\put(180,180){\begin{tabular}{ll}
$e$  &$  {}^{\multiput(0,0)(20,0){4}{\line(10,0){10}}} $ \\[-.7ex] 
$s$  &$  {}^{\put(0,0){\line(70,0){70}}}${}              \\[-.7ex]
$q$  &$  {}^{\multiput(0,0)(10,0){7}{\line(5,0){5}}}   ${}\\[-.7ex]
$m$  &$  {}^{\multiput(0,0)(20,0){3}{\put(0,0){\line(10,0){10}}
         \put(14,0){\line(2,0){2}}}\put(60,0){\line(10,0){10}}}${}
\end{tabular}}
\psfig{file=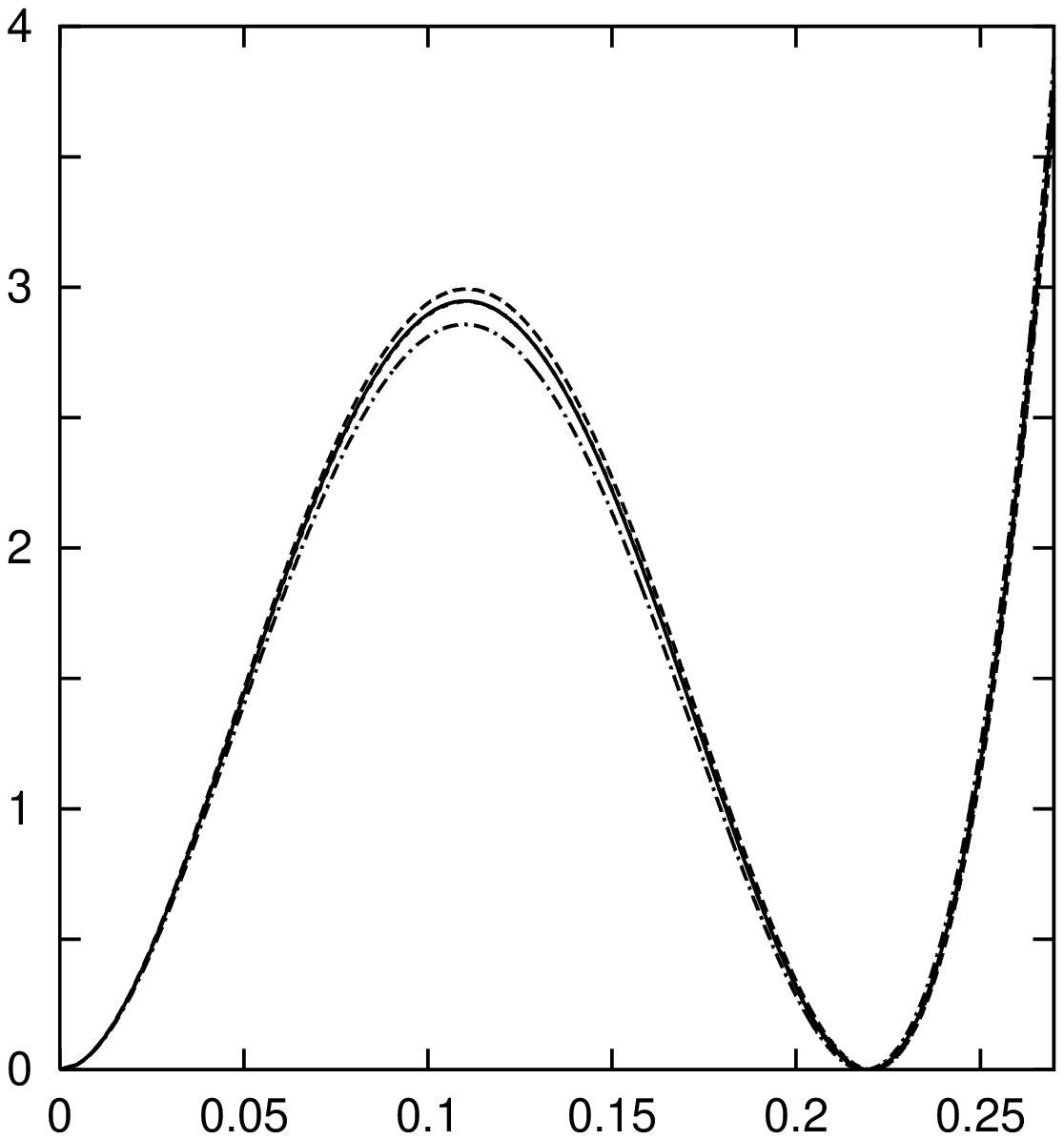,width=480\unitlength}
\hskip-30\unitlength
\psfig{file=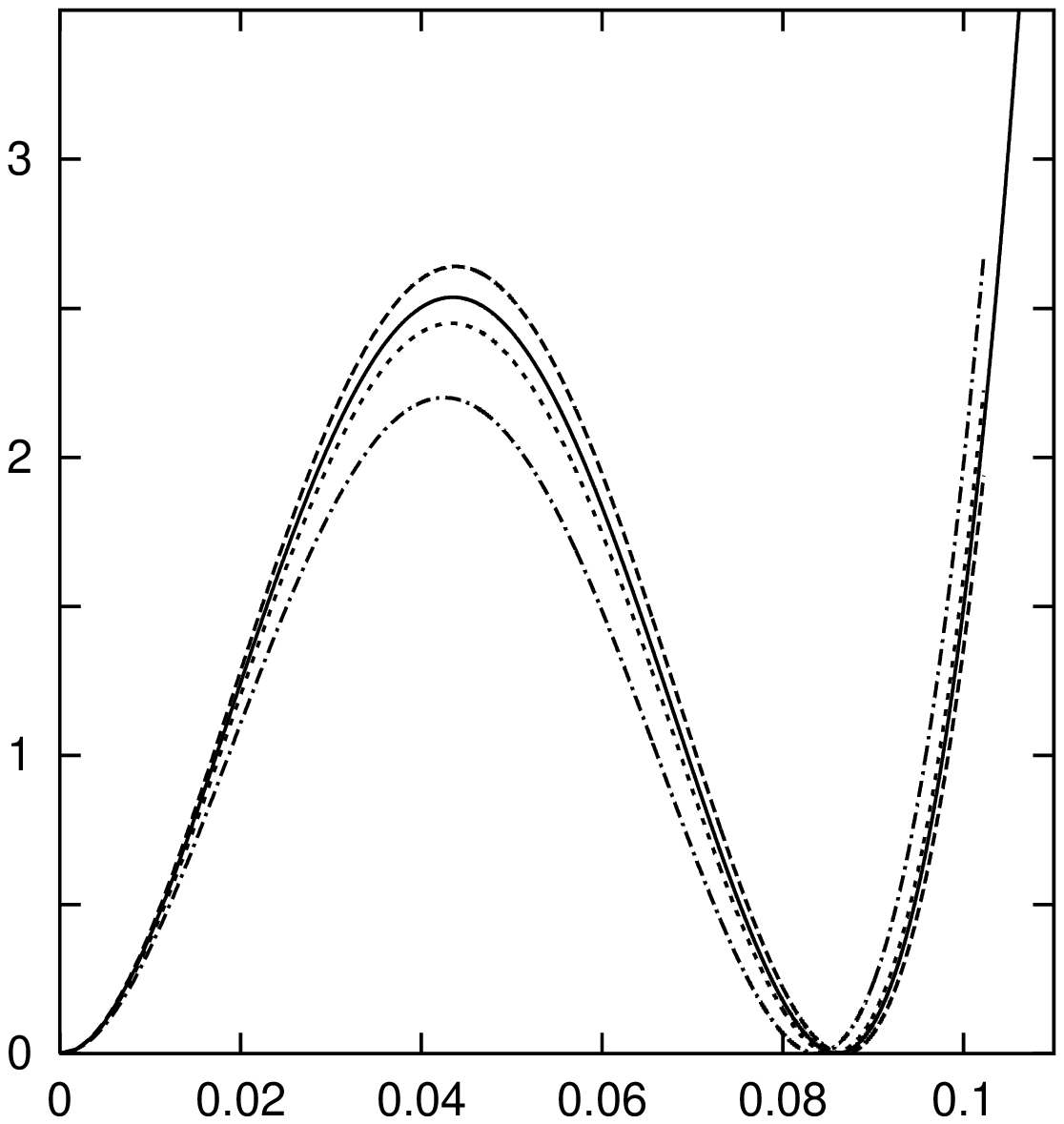,width=480\unitlength}
\end{picture}
\vskip1cm
\begin{minipage}{.94\hsize}
\caption{\label{Ucritscheme} \small The critical potential for 
$m_{\rm H}=38$~GeV (left panel) and $m_{\rm H}=70$~GeV (right panel), 
and different regulator schemes: exponential regulator (e), 
sharp cut-off (s), quartic regulator (q) and mass-like regulator (m).}
\end{minipage} 
\end{center}
\end{figure}

\noindent
Here, we consider the task of computing the critical potential for
coarse grainings other than the sharp cut-off. 
First, we have to obtain the corresponding
fluctuation integrals. The most general expression (for arbitrary
scheme) has been given in Appendix~\ref{AppB}. This expression still 
contains an integral over
momenta to be performed, which is how the scheme dependence enters
into the expression for the fluctuation integral $\Delta_k$. Then, the
criticality conditions \eq{crit} have to be solved to find $T_c$ and
$\rb_0$. The sharp cut-off allowed an analytical computation of
$\Delta_k$, \eq{SharpExpl}, and thus of the functions $F_{1,2}$ in
\eq{F}.
\\[1ex]
Below, in addition to the sharp cut-off, we consider the classes of
power-like regulators
\eq{power} and exponential regulators \eq{exp}. From the 
power-like regulators, we consider the limiting cases $n=1$
({\it i.e.}~a mass-like regulator $R_k=k^2$) and $n=\infty$ 
(the sharp cutoff). As an intermediate case we consider also the 
case $n=2$ ({\it i.e.}~the quartic regulator $R_k=k^4/q^2$). The 
exponential regulators are represented for $n=1$ 
({\it i.e.}~$R_k=q^2/(\exp q^2/k^2 -1)$), and $n=\infty$
(the sharp cutoff). A continuity argument suggests that the critical 
potentials for intermediate values of the coarse graining parameter $n$ 
should appear within those limits set by $n=1, 2$ and $n=\infty$. 
\\[1ex]
No explicit analytical expressions for the
coarse grained free energy have been found in these cases. For the
mass-like and the quartic regulator we used the integrals \eq{Delta-m}
and \eq{Delta-q}, respectively, while \eq{Delta-gen} is used for the
exponential regulator. Then, the problem of solving the criticality
conditions reduces to the optimization of two integral equations.
\\[1ex]
We find that the critical temperature $T_c$ depends very weakly on the
different schemes. Indeed, plotting $T_c$ as a function of the Higgs
field mass we find that the lines corresponding to different schemes
are almost on top of each other, inducing a relative error well below
the $1\%$ level (and thus below the error already present due to other
approximations). A similar situation holds for the  v.e.v., where we
find a relative error below a few percent.
\\[1ex]
In Fig.~\ref{Ucritscheme}, the entire critical potential (in units of
the $4d$ v.e.v.) is displayed for different coarse grainings at
$m_{\rm H}=38$\,GeV (left panel) and at $m_{\rm H}=70$\,GeV (right
panel). The labels $s,\ q,\ m$ and $e$ denote respectively the $s$harp
cut-off, the $q$uartic/$m$ass-like regulator, and the $e$xponential
cut-off from \eq{exp} for $n=1$ and $c=1$. 
\\[1ex]
We first consider $m_{\rm H}=38$\,GeV, and notice that the $s$ and $q$
lines turn out to be on top of each other. Furthermore, it is realised
that the v.e.v.~is nearly independent on the RS, as is the shape of
the potential close to the minima. The main dependence concerns the
local maximum of the critical potential. This dependence will
therefore affect integrated quantities like the surface tension, but
not those related to the v.e.v., like the latent heat. The error for
the surface tension in the present case is about a few percent.
\\[1ex]
For $m_{\rm H}=70$\,GeV the dependence on the scheme is more pronounced
than in the previous case. Furthermore, the v.e.v.~receives\,\,-- for the
mass-like regulator --\,\,a sizeable shift towards smaller values. Again, the
variance is strongest around the maximum of the critical potential,
and dominant in the non-convex region of the critical potential. The 
additional shift in the value of the v.e.v.~entails a corresponding shift
for the outer region of the effective potential, as opposed to the case
for smaller Higgs field mass.
\\[1ex]
It is interesting to make contact with the qualitative considerations
presented at the beginning of this section, and to compare the scheme
dependence observed in Fig.~\ref{Ucritscheme} with the reliability of
the coarse grained potential in its different regions, due to the
approximations employed. Recall that the present computation is based
on neglecting the scalar fluctuations. This approximation is more
reliable for the outer part of the potential than for the non-convex
part of it (more precisely, around a small region of the maximum of
the inner part of the potential). Here, scalar fluctuations cause
ultimately the flattening of the potential in the IR limit. While we
have seen in Sect.~\ref{char-scales} that this approximation is still
reliable for $m_{\rm H}=38$\,GeV, we certainly expect larger
corrections for $m_{\rm H}=70$\,GeV (see the discussion of 
Sects.~\ref{char-scales} and \ref{error}). 
It is quite remarkable that the
scheme dependence indeed seems to reflect the weakness of the
approximation for this region of the potential. Our computation thus
turns the qualitative statement into a quantitative one.
\noindent
Finally, we briefly comment on the different regulators used. It is well-known
that the mass-like regulator is marginal in the sense that it has a 
poor UV behavior which makes its
use for certain applications questionable (a more refined discussion 
has been given in \cite{LP2}). From Fig.~\ref{Ucritscheme}, we learn 
that the critical
potential as obtained for the mass-like regulator deviates the most
from the results for the other regulators employed. 
Considering the class of power-like regulators, we see
from Fig.~\ref{Ucritscheme} that the width between the quartic and the
sharp cut-off limit is significantly smaller than the deviation for
the mass-like regulator. This observation strongly suggests that the
mass-like regulator should be discarded for quantitative
considerations, although it remains, in the present example, a useful 
regulator for studying the main qualitative features of the 
problem.\footnote{This conclusion coincides with those of \cite{LP2} 
based on more formal considerations regarding mass-like regulators.} 
Discarding the mass-like regulator from our discussion, we end up with 
the observation that the error induced through the scheme dependence is 
of the same order of magnitude for algebraic as for exponential regulators. 
For the present case, and at this level of accuracy, no further qualitative 
differences are observed between the exponential regulators \eq{exp} and 
the power-like ones \eq{power} for $n\ge 2$.
\\[1ex]
In summary, we conclude that a quantitative analysis of the scheme
dependence indeed yields non-trivial information regarding the
accuracy of the  approximations or truncations employed, as suggested
by the qualitative  argument presented in Sect.~\ref{quali}. In
addition, we have found some evidence for why a mass-like regulator,
as opposed to exponential or higher order power-like regulators,
should be discarded for accurate quantitative considerations. However,
as the qualitative features are still well described by a mass-like
regulator, and as the quantitative deviation is not too big, this also
suggests that a mass term regulator could be very useful for an
error estimate.\footnote{An error estimate based on the mass-like
regulator is rather conservative as it seems to overestimate the
scheme dependence.} Typically, analytical computations are largely
simplified for mass-like regulators, allowing for a simple cross-check
of the results.

\section{Summary and outlook}\label{discussion}
\noindent
We have studied in detail the first-order phase transition of Abelian
Higgs models in 3+1 dimensions at finite temperature. Properties of 
the transition are determined by the underlying fixed point structure of 
the $3d$ theory such as the cross-over of the Abelian charge from the Gaussian 
to the Abelian fixed point.
We computed
all physical observables at the phase transition, the phase diagram
in the domain of first-order transitions
and the tri-critical point. The analysis has
been restricted to the region of parameter space where the dimensional 
reduction scenario applies and a perturbative matching of the $4d$ 
parameters to the corresponding $3d$ ones is possible. The main 
contribution to the
free energy (and thus to the physical observables at criticality) stem
from the remaining effective $3d$ running for which we have used a Wilsonian
renormalisation group to leading order in the derivative
expansion, neglecting the scalar, but not the gauge field anomalous 
dimension. The latter is related to the non-trivial running of the 
Abelian gauge coupling, which is described by an effective fixed point. 
While this fixed point
is well understood in the large\,-$N$ limit where the tri-critical fixed
point is known, its precise form is not yet established for the
relevant case of $N=1$. We therefore studied the parametric dependence
of physical observables on the fixed point value. A quantitative discussion
of the relevant physical scales, which are easily accessible within
a Wilsonian framework, has also been given.   
\\[1ex]
The main effect on physical observables due to the presence of a
non-trivial fixed point depends on the ratio between the cross-over
scale $k_{\rm cr}$ (which defines the cross-over to the Abelian fixed point) 
and the typical scales characterising the first-order phase transition (like 
the discontinuity scale $k_{\rm dis}$, or $k_{\rm stable}$).
For $k_{\rm cr}$ small as compared to $k_{\rm stable}$ the observed
dependence is weak. The sizeable deviations from the perturbative
$\eb2(k)\approx\eb2(\La)$\,-\,behaviour only set in at very small 
scales below $k_{\rm stable}$ and are no longer
relevant for the phase transition itself in this situation.
The main effects are restricted to alterations in the far
infra-red region, like the details of the  
flattening of the inner part of the potential. 
On the other hand, a strong dependence emerges for $k_{\rm cr}$ larger
than $k_{\rm stable}$.
\begin{figure}
\begin{center}
\unitlength0.001\hsize
\begin{picture}(700,600)
\put(300,80){{\Large $m_{\rm H}$} [{GeV}]}
\put(170,535){\mbox{{\large  $T/T_{\rm ref}:$}}}
\put(170,470){\mbox{{\large  $\rb/\rb_{\rm ref}:$}}}
\put(350,535){{\footnotesize $2\le e_\star\le\infty$}}
\put(350,495){{\footnotesize $6\le e_\star\le \infty$}}
\put(350,471){{\footnotesize $4\le e_\star\le 6$}}
\put(350,447){{\footnotesize $2\le e_\star\le 4$}}
\hskip100\unitlength
\psfig{file=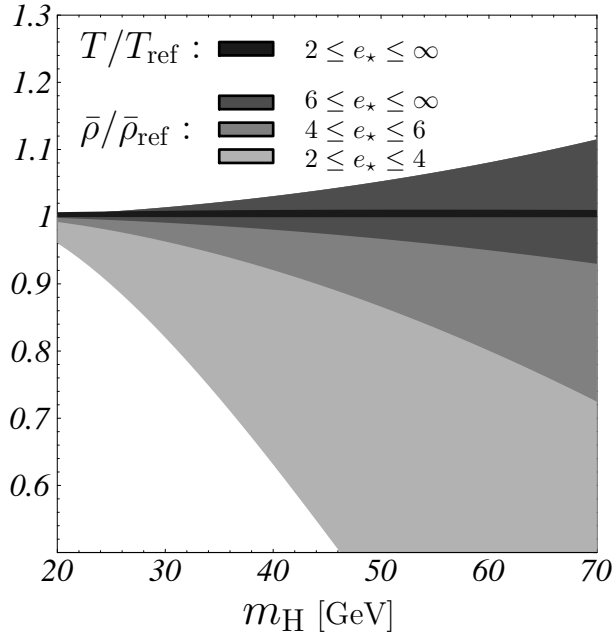,width=500\unitlength}
\end{picture}
\vskip-1cm
\begin{minipage}{.8\hsize}
\caption{\label{varianceEs} \small The relative variation of $T_c$ and $\rb_0$ 
with the effective Abelian fixed point. Here, the regions $T/T_{\rm ref}$ and
$\rb/\rb_{\rm ref}$ compare the critical temperature and the v.e.v.~as
a function of $e_\star$. The reference values are obtained for
$e_\star=\sqrt{6}\pi\approx 7.7$.}
\end{minipage} 
\end{center}
\end{figure}
\noindent
Most of our results for the physical observables can be summarised as in
Fig.~\ref{varianceEs}. Here, the reference values $T_{\rm ref}$ and 
$\rb_{\rm ref}$ are given for $e_\star=\sqrt{6}\pi$ (which corresponds 
roughly to $k_{\rm cr}\approx k_{\rm stable}$), and for the sharp cut-off
regulator. In the present approximation, the critical temperature 
is insensitive to the
running gauge coupling. On the other hand, the v.e.v.~appears to be
quite sensitive to the actual fixed point value, in particular for
larger Higgs field mass. The phase transition weakens significantly
for small fixed point values. The reason is that the gauge coupling is
decreasing strongly for small fixed point values at scales larger than
the scale where the critical potential  reaches its degenerate shape,
that is above the scale of decoupling. These results compare well with
perturbation theory, except for very large or very small values for the 
Abelian fixed point.  Corrections due to the non-trivial
scaling of $\bar e^2(k)$ remain below $10\%$  for $\es2 >6\pi^2$ and
$m_{\rm H}$ below 70\,GeV, but do grow large as soon as $\es2$ is a
below $6\pi^2$. We conclude that $\es2 \approx 6\pi^2$ 
is a good leading order approximation for small Higgs field mass
as higher order corrections are small. For $m_{\rm H}=30$\,GeV, we also 
compared the value for the critical temperature with lattice simulations 
and found agreement below $4\%$.
The sensitivity 
on $\es2 < 6\pi^2$ for larger Higgs mass, in turn, requires a better 
determination of the fixed point in this domain. This concerns 
in particular physical observables like the critical exponents at the 
endpoint of the line of first order phase transitions.
\\[1ex]
For generic regulator function the free energy in the type-I regime 
has been given as an integral (one remaining integration). For the 
case of a sharp cut-off regulator, we obtained an explicit
analytical solution for the free energy, 
given the non-trivial scale
dependence of the Abelian charge. In the present article, we evaluated
all relevant quantities for initial conditions obtained from a
perturbative dimensional reduction scenario relevant for a high 
temperature (cosmological) phase transition. 
\\[1ex] 
The explicit result for the
effective potential can also be of use for applications to the 
superconducting phase transition, or for the nematic to smectic-A phase
transition in certain liquid crystals. The main change would concern the
initial potential for the effective $3d$ flow of the potential, and
the numerical value of the Abelian charge at that scale. These changes
affect in particular the ratio $k_{\rm cr}/k_{\rm stable}$, and therefore
the above discussion, as both scales depend in a qualitatively different 
manner on $e^2(\La)$ and $U_\La$.
\\[1ex]
In addition, we studied the dependence of our
results on the coarse graining procedure employed. 
We have seen that the physical observables do
depend only very weakly on the coarse graining. This is encouraging, as
a strong dependence would have cast serious doubts on the
approximations used. Furthermore, we employed a variety of 
qualitatively different
coarse grainings ranging from the mass-like and other polynomial
regulators over  exponential ones to the sharp cut-off regulator. Therefore,
our
result can be seen as an important consistency check of the
method. The weak variation w.r.t.~the coarse graining which is to be
interpreted as an `error bar' for the observables, is smaller or of
about the same size as the error expected from higher order operators
for the coarse grainings studied. This `error bar' would vanish only if
no truncation to the effective action would have to be performed.  
We also observed
an intimate relationship between the truncation of the effective
action, and the error bar introduced through the scheme
dependence. More precisely, it is observed that the scheme dependence
is largest in regions where a similarly large effect due to the
neglecting of the scalar fluctuations in the non-convex region of the
potential is expected. While this result is not entirely unexpected,
a quantitative evidence for it has never been presented before. It would 
be useful if further quantitative results in this direction could be 
established. This concerns in particular the cross-dependences between
an optimal coarse-graining that minimizes the scheme dependence, and an 
optimized convergence of systematic
truncations and approximations \cite{Scheme}.
\\[1ex]
An important open question for future work concerns the
precise IR behaviour of the Abelian charge. This, of course, is an
intrinsic problem of the $3d$ theory. As argued, our current
understanding is mainly limited due to an insufficient understanding
of the field and/or momentum dependence of the Abelian charge. It might be
fruitful to consider alternatively a thermal renormalisation group
to improve the situation \cite{Litim98}. At the same time, the
inclusion of higher order corrections due to scalar fluctuations will
also become important\,\,-- close to the critical points --\,\,for a
reliable determination of critical exponents and other universal 
quantities. It would also be interesting to consider the $SU(2)$-Higgs 
theory, where a non-trivial endpoint of the line of first-order phase 
transitions has been established recently. A field theoretical 
understanding of this endpoint is still missing, and a derivation of 
the related critical indices from field theory would be desirable. 
Again, one expects that the IR behaviour of the gauge coupling, in 
competition with the scalar fluctuations, is responsible for the 
existence of the endpoint.

\begin{appendix}
\renewcommand{\theequation}{A.\arabic{equation}}
\setcounter{equation}{0}

\section{RS dependence and threshold functions}\label{AppA}
\noindent
The solution of the flow equation (and the related physical observables) 
can be written as momentum integrals over a measure, which depends on the 
precise implementation of the coarse graining. We employ the notation of
\cite{Litim97}, where a scheme dependent measure has been given
(in $d$ dimensions) as 
\beq \label{I}
{I}_r[f] = -\0d2 \int_0^\infty dy  \0{r'(y)}{(1+r(y))^{1+d/2}}  f(y)
\eeq
for momentum-dependent functions $f(y)$, where $y=q^2/k^2$, and $q$ is the 
loop momenta. As a consequence 
of the conditions \eq{ConditionsRk} on the regularisation function $r(y)$ 
it follows that the momentum measure $-r'(y)/(1+r)^{1+d/2}$ is peaked. 
The measure is normalised to one,  
\beq \label{Inorm}
 {I}_r[1] = 1.
\eeq
This implies that $I_r[f]$ depends on the coarse graining as soon 
as $f$ displays a non-trivial dependence on momenta. 
\\[1ex]
As an example, let's consider the threshold functions $\ell^d_n(\om)$, 
defined as
\beq
\ell^d_n(\om)=-\left(\de_{n,0}+n\right)\int^\infty_0 dy
\0{r'(y)y^{1+d/2}}{[y(1+r)+\om]^{n+1}} \ .
\eeq
They are related to the above measure through
\beq
\ell^d_n(\om)=\02d \left(\de_{n,0}+n\right)
{I}_r\left[\0{P^{d+2}}{(P^2+\om)^{n+1}} \right]\ .
\eeq
Here, we also introduced the dimensionless effective (regularised) 
inverse propagator
\beq \label{P}
P^2(y) = y + y\ r(y)  \ .
\eeq
The threshold functions can always be expanded as a Taylor series
in powers of $\om$. Let 
us define the corresponding RS dependent expansion coefficients
\beq\label{ak}
a_k = {I}_r[P^{-k}]\ ,
\eeq
which are the $k^{th}$ moments of $1/P$ w.r.t.~the measure ${I}_r$. 
These coefficients appear in the computation of the endpoint of the 
critical line \eq{endpoint},
which is proportional to the coefficient $a_1$. For a power-like 
regulator $r(y)=y^{-n}$ [see \Eq{power}] we find for arbitrary dimension $d$
\beq
a_k =\s0d2\ \Gamma[1+
\s0k{2n}]\0{\Gamma[\s0d2+\s0k2(1-\s01n)]}{\Gamma[1+\s0d2+\s0k2]} \ .
\eeq
A more detailed discussion of these coefficients and a related discussion 
of the convergence of
amplitude expansions and optimised coarse-graining parameters
is given in \cite{Scheme}.

\renewcommand{\theequation}{B.\arabic{equation}}
\setcounter{equation}{0}


\section{The fluctuation integral}\label{AppB}
\noindent
The fluctuation integral reads
\beq\label{deltadef2}
\De_{k}(\rb)=
-\0{1}{2\pi^2}\int^\La_kd\bar k \int^\infty_0dy\ \0{\bar k^2}{P^2}\ 
\0{2\es2\rb\ r'(y)\ y^{5/2}}{2\es2\rb +P^2\bar k(1+\bar k/\kcr)} 
\eeq
Note that we have normalised $\De(0)=0$ in the above definition. The 
remaining integrals in \eq{deltadef2} can be solved in different ways, 
either first performing the momentum integration or the scale integration. 
Integrating first w.r.t.~$\bar k$ yields (for the notation see Appendix A)
\beq\label{Delta-gen}
\De_{k}(\rb)={I}_r\left[{\cal U}(\rb,P)\right]
\eeq
where
\bea \label{Ugen2}
{3 \pi^2\ \cal U}(\rb,P)
&=&\nonumber 
2 \es2\rb\int^\La_k d\bar k \0{P^3 \bar k^2}{P^2\bar k(1+\bar
k/\kcr)+2\es2\rb}  \\
&=&-2\es2\rb\ P\ \kcr (k-\La)
   +{\es2\rb}\ P\ \kcr^2\ln \left(
   \0{2\es2\rb/P^2+k+k^2/\kcr}{2\es2\rb/P^2+\La+\La^2/\kcr} 
   \right)
   \nonumber \\ \label{Ugen3} &&
   +2 \es2\rb\ P\ \kcr(4\es2\rb/P^2-\kcr)\ 
    G_{k,\Lambda}(1-8\es2\rb/P^2\kcr)  \ , 
\eea
with $I_r$ defined in \eq{I} and $P(y)$ in \eq{P}. 
The function  $G(\Omega)$ reads
\beq\label{G}
G_{k,\Lambda}(\Omega)=\left\{ 
\begin{array}{c@{\quad \mbox{for}\quad}l}
\displaystyle \01{2 \sqrt{\Omega}} \ln\left( 
\0{1+2 k/\kcr-\sqrt{\Omega}}{1+2 k/\kcr+\sqrt{\Omega}}~
\0{1+2\La/\kcr+\sqrt{\Omega}}{1+2 \La/\kcr-\sqrt{\Omega}}\right)  
& \Omega > 0 \\[4ex]
\displaystyle \0{-1}{\sqrt{-\Omega}}\left[\arctan\left(
\0{{\sqrt{-\Omega}}}{1+2 k/\kcr}\right)-
\arctan\left(\0{{\sqrt{-\Omega}}}{1+2 \La /\kcr}\right)\right] 
& \Omega < 0 \\[4ex]
\displaystyle \0{2\kcr (k-\La)}{(\kcr +2 k)(\kcr+2 \La)}&\Omega = 0~. 
\end{array} \right.
\eeq
For a sharp cut-off regulator \eq{sharp}, the remaining momentum 
integration can be performed analytically to give
\bea\label{SharpExpl}
2\pi^2 \De_k^{(s)} &=&\s013 \La^3
\ln\left(1+\0{2\es2\rb\kcr}{\La(\La+\kcr)}\right) -
\s013 k^3 \ln\left(1+\0{2\es2\rb\kcr}{k(k+\kcr)}\right) -
\s013 \kcr^3 \ln\left(\0{\La+\kcr}{k+\kcr}\right) \nonumber \\[1ex]
&&+\s016(\kcr^3-6\es2\rb\kcr^2)\ln\left(\0{\La(\La+\kcr)+2\es2\rb\kcr
}{k(k+\kcr)+2\es2\rb\kcr }\right) +\s043\es2\rb\kcr(\La-k)\nonumber \\[2ex]
&&-\s013\kcr(-16\es4\rb^2
+10\es2\rb\kcr-\kcr^2)\ G_{k,\Lambda}\left(1-{8\es2\rb}/{\kcr}\right) \ .
\eea
We have normalised $\De(\rb)$ such that $\De(0)=0$.
\\[1ex]
On the other hand, performing first the scheme dependent momentum 
integration leaves us with the following remaining integrals,
\bea \label{Delta-s}
\Delta^{(s)}(\rb)&=&\01{2\pi^2} \int_k^\La d\bar k\ \bar k^2
\ln\left({1+\0{2\es2\rb}{\bar k(1+\bar k/\kcr)}}\right)
\\ \label{Delta-m}
\Delta^{(m)}(\rb)&=&\01{2\pi} \int_k^\La d\bar k\ \bar k^2\left(
\sqrt{1+\0{2\es2\rb}{\bar k(1+\bar k/\kcr)}}-1\right)
\\ \label{Delta-q}
\Delta^{(q)}(\rb)&=&\01{\sqrt{2}\pi} \int_k^\La d\bar k\ \bar
k^2 \left(1-\left(1+\0{\es2\rb}{\bar k(1+\bar k/\kcr)}\right)^{-1/2}\right)
\eea
Here, the indices refer to the sharp ($s$), the mass-like ($m$) and the
quartic ($q$) cut-off function, as defined in Sect.~\ref{schemes}. 

\renewcommand{\theequation}{C.\arabic{equation}}
\setcounter{equation}{0}


\section{Including scalar fluctuations}\label{AppC}
\noindent
In order to obtain an estimate of the effect of the scalar fluctuations we 
will solve \Eq{flowpot} with $U_k$ on the r.h.s.~replaced by $U_\La$, 
for a sharp cut-off regulator. The flow equation becomes
\beq\label{flowpot2}
4 \pi^2\0{dU_k(\rb)}{k^2dk}=\el 30
\left(\0{m^2_{1,\La}(\rb)}{k^2}\right) +\el 30
\left(\0{m^2_{2,\La}(\rb)}{k^2}\right) + 
2\el 30 \left(\0{2\eb2_3(k)\rb}{k^2}\right)\ .
\eeq
with the masses $m_i^2$   given through
\beq\label{m12}
m_1^2(\rb)=m_{\rm R}^2 +  \bar \la_{\rm R} \rb,\quad 
m_2^2(\rb)=m_{\rm R}^2 + 3\bar \la_{\rm R} \rb\ .
\eeq
We introduce the functions 
\bea
K(m^2)&=&-\0{1}{4\pi^2}\int^{k}_\La dy y^2
\ln\left(1+\0{m^2}{y^2}\right)\nonumber \\
&=&-\01{12\pi^2}\left[2 m^2 k- 2m^3 {\rm arctan}\left(\0km\right) +k^3
\ln\left(1+\0{m^2}{k^2}\right)-(k\leftrightarrow \La)\right]\\
J_i(\rb)&=&K[m_i^2(\rb)]-K[m_i^2(0)]\ .
\eea
The solution to the flow \eq{flowpot2} then obtains, using also $\De^{(s)}$ 
from \eq{SharpExpl}, as
\beq\label{SharpExpl2}
U_k(\rb)=U_\La(\rb)+\De^{(s)}(\rb)+J_1(\rb)+J_2(\rb)\ .
\eeq
The effect of the additional terms on the shape of the critical potential 
is about a few percent, increasing towards higher values for $m_{\rm H}$.

\end{appendix}


\begin{thebibliography}{99}
\def\BOOK#1#2#3#4{#1, {\sc } #3, #4}
\def\PRA#1#2#3#4#5{ #1,\,\,{\it }\,Phys.\,Rev.\,{\bf A#3}\,(19#4)\,#5}
\def\PRB#1#2#3#4#5{#1,{\it }\,Phys.\,Rev.\,{\bf B#3}\,(19#4)\,#5}
\def\PRL#1#2#3#4#5{#1,\,\,{\it }\,Phys.\,Rev.\,Lett.\,{\bf #3} (19#4) #5}
\def\PRC#1#2#3#4#5{#1,\,\,{\it }\,Phys.\,Rev.\,{\bf C#3}\,(19#4)\,#5}
\def\PRD#1#2#3#4#5{#1,{\it }\,Phys.\,Rev.\,{\bf D#3}\,(19#4)\,#5}
\def\PRE#1#2#3#4#5{#1,\,\,{\it }\,Phys.\,Rev.\,{\bf E\,#3}\,(19#4)\,#5}
\def\PRep#1#2#3#4#5{#1,{\it }\,Phys.\,Rep.\,{\bf  #3}\,(19#4)\,#5}
\def\NPB#1#2#3#4#5{#1,{\it }\,Nucl.\,Phys.\,{\bf B#3}\,(19#4)\,#5}
\def\PLB#1#2#3#4#5{#1,{\it }\,Phys.\,Lett.\,{\bf B#3}\,(19#4)\,#5}
\def\PTP#1#2#3#4#5{#1,\,\,{\it }\,Prog.\,Theor.\,Phys.\,{\bf B#3}\,(19#4)\,#5}
\def\SSC#1#2#3#4#5{#1,\,\,{\it }\,Solid\,State\,Comm.\,{\bf #3} (19#4) #5}
\def\EPL#1#2#3#4#5{#1,\,\,{\it }\,Europhys. Lett.~{\bf #3}\,(19#4)\,#5}
\def\JCP#1#2#3#4#5{#1,\,\,{\it }\,J.\,Phys.\,(Paris) {\bf  #3} (19#4) #5}
\def\JPA#1#2#3#4#5{#1,\,\,{\it }\,J.\,Phys.\,{\bf A#3}\,(19#4)\,#5}
\def\JPB#1#2#3#4#5{#1,\,\,{\it }\,J.\,Phys.\,{\bf B#3}\,(19#4)\,#5}
\def\JPC#1#2#3#4#5{#1,\,\,{\it }\,J.\,Phys.\,{\bf C#3}\,(19#4)\,#5}
\def\ZPC#1#2#3#4#5{#1,{\it }\,Z.\,Phys.\,{\bf C#3}\,(19#4)\,#5}
\def\JETP#1#2#3#4#5{#1,\,\,{\it }\,Soviet\,Physics\,JETP\,Lett.\,{\bf #3}\,(19#4)\,#5}
\def\MPLA#1#2#3#4#5{#1,{\it }\,Mod.\,Phys.\,Lett.\,{\bf A#3}\,(19#4)\,#5}
\def\PA#1#2#3#4#5{#1,\,\,{\it }\,Physica\,{\bf A#3}\,(19#4)\,#5}
\def\PS#1#2#3#4#5{#1,\,\,{\it }\,Physics\.{\bf #3}\,(19#4)\,#5}
\def\AP#1#2#3#4#5{#1,\,\,{\it }\,Ann.\,Phys.\,{\bf #3}\,(19#4)\,#5}
\def\IJMPA#1#2#3#4#5{#1,{\it }\,Int.\,J.\,Mod.\,Phys.\,{\bf A#3}\,(19#4)\,#5}
\def\LNC#1#2#3#4#5{#1,\,\,{\it }\,Lett.\,Nuevo\,Cimento\,{\bf #3}\,(19#4)\,#5}
\def\PPR#1#2#3{#1,\,\,{\it }\,Preprint\,#3}
\def\and#1#2#3{{\bf #1}\,(19#2)\,#3}

\bibitem{LuMa} 
   \PRL{B.I.\,Halperin, T.C.\,Lubensky and S.\,Ma}{}{32}{74}{292}; \\
   H.\,Kleinert, {\it Gauge fields in condensed matter} 
   (World Scientific, 1989).

\bibitem{deGennes}P.G.\,de Gennes and J.\,Prost, 
   {\it The Physics of Liquid Crystals} (Cambridge Univ.~Press, 1993).

\bibitem{Wegner73}
   \PRA{F.J.\,Wegner and A.\,Houghton}{Renormalization Group
       Equation for Critical Phenomena,}{8}{73}{401}; \\
   \PRep{K.G.\,Wilson and I.G.\,Kogut}{}{12}{74}{75}. 

\bibitem{Polchinski84}
   \NPB{J.\,Polchinski}{Renormalization and
   Effective Theories,}{231}{84}{269}. 

\bibitem{Wetterich91}\,\NPB{C.\,Wetterich}{}{352}{91}{529};
\,Phys.\,Lett.\,\and{B301}{93}{90};\,Z.Phys.\,\and{C57}{93}{451}. 

\bibitem{ReuterWetterich93}
   \NPB{M.\,Reuter and C.\,Wetterich}{}{391}{93}{147}; 
   \and{B408}{93}{91}; 
   \and{B427}{94}{291}.

\bibitem{OSF93}
   \MPLA{D.\,O'Connor,\,\,C.R. Stephens\,\,and\,\,F.\,Freire}{}{8}{93}{1779}.

\bibitem{Tetradis93}
   \NPB{N.\,Tetradis and C.\,Wetterich}{}{398}{93}{659}.   

\bibitem{BFLLW}
   \PRB{B.\,Bergerhoff,\,\,F.\,Freire,\,\,D.F.\,Litim,\,\,S.\,Lola\,\,and\,\,C.\,Wetterich}{Phase Diagram of Superconductors from Non Perturbative
   Flow Equations}{53}{96}{5734}, [{hep-ph/9503334}].

\bibitem{BLLW}
   \IJMPA{B.\,\,Bergerhoff,\,\,D.F.\,Litim,\,\,S.\,Lola\,\,and\,\,C.\,Wetterich}{Phase Transition of $N$-Component Superconductors}{11}{96}{4273},\newline
   [{cond-mat/9502039}].

\bibitem{TL96}
   \NPB{N.\,Tetradis and D.F.\,Litim}{Analytical Solutions of Exact 
   Renormalisation Group Equations}{464}{96}{492}, [{hep-th/9512073}].

\bibitem{Tetradis97}
   \NPB{N.\,Tetradis}{}{488}{97}{92}, [{hep-ph/9608272}].

\bibitem{Litim97}
   \PLB{D.F.\,Litim}{Scheme Independence and First Order
   Phase Transitions from Exact Renormalization Group
   Equations}{393}{97}{103}, [{hep-th/9609040}].

\bibitem{PT1}
   \NPB{W.\,Buchm\"uller, T.\,Helbig and D.\,Waliser}{}{407}{93}{387}.

\bibitem{Hebecker93}
   \ZPC{A.\,Hebecker}{Finite Temperature Effective
   Potential for the Abelian Higgs Model to the Order
   $\e4,\la^2$}{60}{93}{271}, [{hep-ph/9307268}].

\bibitem{GI}
   \NPB{W.\,Buchm\"uller, Z.\,Fodor and A.\,Hebecker}{}{447}{95}{317}, 
   [hep-ph/9502321].
 
\bibitem{Dimopoulos:1998cz}
P.~Dimopoulos, K.~Farakos and G.~Koutsoumbas,
Eur.\ Phys.\ J.\ C {\bf 1} (1998) 711
[hep-lat/9703004].

\bibitem{lattice}
   K.\,Kajantie,\,\,M.\,Karjalainen,\,\,M.\,Laine\,\,and\,\,J.\,Peisa,\,\,Phys.\,Rev.\,{\bf B57}\,(1998)\,3011, [{cond-mat/9704056}];
   Nucl.\,Phys.\,{\bf B520} (1998) 345, [{hep-lat/9711048}].

\bibitem{Ellwanger}
   \PLB{U.\,Ellwanger}{}{335}{94}{364}, [{hep-th/9402077}].

\bibitem{FW} 
   F.\,Freire and C.\,Wetterich, Phys.\,Lett.\,{\bf B380} (1996) 337,
   [{hep-th/9601081}]. 

\bibitem{LP1}
   \PLB{D.F.\,Litim\,\,and\,\,J.M.\,Pawlowski}{}{435}{98}{181}, [{hep-th/9802064}];\\
   Nucl.\,Phys.\,{\bf B74} (PS) {(1999)} {329}, [{hep-th/9809023}];
   {\bf B74} (PS) {(1999)} {325}, [{hep-th/9809020}]. 

\bibitem{LP2}
   D.F.\,Litim and J.M.\,Pawlowski, {\it On gauge invariant 
   Wilsonian flows}, [{hep-th/9901063}].

\bibitem{Morris94}
   \PLB{T.\,Morris}{}{329}{94}{241}, [{hep-ph/9403340}].

\bibitem{Litim98}
   {D.F.\,Litim}, {\it Wilsonian flow equations and thermal 
   field theory}, [{hep-ph/9811272}].

\bibitem{Pietroni}
   \NPB{M.\,d'Attanasio and M.\,Pietroni}{}{498}{97}{443},
   [{hep-th/9611038}].

\bibitem{Kajantie96a}
   \NPB{K.\,Kajantie,\,\,M.\,Laine,\,\,K.\,Rummukainen\,\,and\,\,M.\,Shaposhnikov}{}{458}{96}{90}, [{hep-ph/9508379}]. 

\bibitem{Litim94a}
   \MPLA{D.F.\,Litim,\,C.\,Wetterich\,and\,N.\,Tetradis}{Non-Perturbative
   Analysis of the Coleman-Weinberg Phase Transition}{12}{97}{2287},\,
   [{hep-ph/9407267}].

\bibitem{ColemanWeinberg73}
   \PRD{S.\,Coleman and E.\,Weinberg}{}{7}{73}{1888}.

\bibitem{Tetradis92}
   \NPB{N.\,Tetradis and C.\,Wetterich}{flattening}{383}{92}{197}.

\bibitem{K94}
   K.\,Farakos,\,\,K.\,Kajantie,\,\,K.\,Rummukainen\,\,and\,\,M.\,Shaposhnikov,\,\,Nucl.\,Phys.\,{\bf B425}\,(1994)\,67, [{hep-ph/9404201}]. 

 
\bibitem{tunneling}
   \NPB{A.\,Strumia and N.\,Tetradis}{}{554}{99}{697}, [{\tt hep-ph/9904246}].

\bibitem{Arnold1}
   P.\,Arnold and L.\,Yaffe, Phys.\,Rev.\,{\bf D49} (1994) 3003, [{
   hep-ph/9312221}].

\bibitem{Arnold2}
   P.\,Arnold and D.\,Wright, Phys.\,Rev.\,{\bf D55} (1997) 6274, [{
   hep-ph/9610226}].

\bibitem{Scheme}
   D.F.\,Litim, Phys.\,Lett.\,  {\bf B486} (2000) 92,
   [{hep-th/0005245}]; hep-th/0103195.
\end{thebibliography}
\end{document}